\documentclass[twocolumn,prb,amsmath]{revtex4}
\usepackage{graphicx}

\newcommand{\ket}[1]{{\vert #1 \rangle}}
\newcommand{\bra}[1]{{\langle  #1 \vert}}
\newcommand{\braket}[2]{{\langle  #1 \vert #2 \rangle}}
\newcommand{\uA}{\underline{\mathrm{A}}}
\newcommand{\ua}{\uparrow}
\newcommand{\da}{\downarrow}

\begin{document}

\author{Iztok Pi\v{z}orn}
\affiliation{Theoretische Physik, ETH Zurich, CH-8093 Z\"{u}rich, Switzerland}

\author{Frank Verstraete}
\affiliation{Vienna Center for Quantum Science, University of Vienna, A-1090 Wien, Austria}
\affiliation{Department of Physics and Astronomy, Ghent University, B-9000 Ghent, Belgium}

\author{Robert M. Konik}
\affiliation{CMPMS Dept., Bldg. 734, Brookhaven National Laboratory, Upton, NY 11973-5000, USA}

\title{Tree tensor networks and entanglement spectra}

\begin{abstract}
A tree tensor network variational method is proposed to simulate quantum many-body systems with global symmetries where the optimization is reduced to individual charge configurations.
A computational scheme is presented, how to extract the entanglement spectra in a bipartite splitting of a loopless tensor network across multiple links of the network, by constructing a matrix product operator for the reduced density operator and simulating its eigenstates.
The entanglement spectra of $2 \times L$, $3\times L$ and $4\times L$ with either open or periodic boundary conditions on the rungs are studied using the presented methods, where it is found that the 
entanglement spectrum depends not only on the subsystem but also on the boundaries between the subsystems.
\end{abstract}

\maketitle

\section{Introduction}
The entanglement entropy, a distinct property of quantum systems, is the most valuable resource in quantum computation and the main object of interest in the field of quantum information.\cite{nielsen} 
 Low degree of entanglement of quantum states at zero temperature is featured in efficient description of quantum systems by approximate computational methods such as the Density Matrix Renormalization Group (DMRG),\cite{whiteprl92,schollwoeck,daley} methods based on matrix product states \cite{ostlund,vidal1,verstraetempo,cmps,znidaric,banuls} or generalized tensor networks.\cite{peps,terg,mera,murg} 
In condensed matter physics, a connection was made between the entanglement entropy
and quantum critical phenomena where it was found \cite{osterloh,rico} that quantum critical systems are characterized by a logarithmic violation of the area law \cite{arealaw} with a central charge corresponding to the underlying conformal field theory.\cite{cardy}

The entanglement entropy, as the logarithmic sum of the eigenvalues of the reduced density matrix, $\rho_{\rm reduced}$ 
resulting from a partition of a quantum system, does not capture all the information available in $\rho_{\rm reduced}$.  Its full spectrum had
been previously studied for intution into the operation of the DMRG algorithm.\cite{peschel1,peschel2,okunishi}
More recently it was discovered\cite{haldane} that the spectrum, not of $\rho_{\rm reduced}$, but of the related operator $\log \rho_{\rm reduced}$,
provided insight into the topological nature of the quantum state from which $\rho_{\rm reduced}$ was derived.  In Ref.~\onlinecite{haldane} it was
demonstrated that the spectrum of $\log \rho_{\rm reduced}$ arising from a $\nu=1/3$ quantum Hall state matched that of the compactified bosonic
theory expected to described the $\nu=1/3$ edge state.  This correspondence in systems with topological order has now been extensively
elaborated on in quantum Hall systems.\cite{regnault,zozulya,lauchli1,thomale,sterdyniak,thomale1,chandran,sterdyniak1,qi}
This flurry of work prompted exploration of other systems, systems which were not necessarily topological.  Entanglement
spectra was studied for insight into the behaviour of critical and non-critical one dimensional systems,\cite{lefevre,pollman,thomale2,albaprl} for the detection/reflection of topological order in one dimensional spin chains, \cite{pollman1} and two dimensional non-topological systems.\cite{james,lauchli2}

Of particular relevance for the work herein, there have been a number of studies of the entanglement spectra of quasi-one-dimensional systems
such as spin ladders\cite{poilblanc,lauchli, peschel3} where a lengthwise partition of the system was considered.  So in the case of a ladder geometry,
a partition of the system cutting the rungs of the ladder was studied.  Remarkably for such partitions the entanglement spectra reflected
the true spectra of the partitioned subsystem.  So in the case of Heisenberg spin ladders divided into two spin chains, the entanglement spectra
appeared to share characteristics of the spinon spectrum of a Heisenberg spin chain,\cite{poilblanc} in particular the entanglement spectra followed
the des Cloizeaux-Pearson lower spinon boundary.\cite{cloizeaux}  This observation was sharpened in\cite{lauchli} where is was shown
for ladders with weak spin-spin couplings along the leg, the entanglement Hamiltonian was exactly that of the Heisenberg spin chain.
One of the aims of this paper is to study the entanglement spectra of multi-legged ladders.

To achieve this goal we must first surmount the problem that it remains computationally expensive to extract the
entanglement spectra for nontrivial geometries.  We can of course always have recourse to exact diagonalization but this limits
us to the study of relatively small systems. While it is trivial to extract the entanglement spectrum in a bipartite splitting of quantum chains using linear tensor network methods such as DMRG where the eigenvalues of the reduced density operator are inherent to the computational scheme, 
the entanglement spectrum in a splitting of the systems where several links in the tensor network are broken poses an exponentially difficult problem.
And this supposes we are even able to describe accurately the ground state of the model of interest.  If the model is defined on the lattice and if one
spatial dimension is sufficiently small, the ground states of such systems are again well approximated by DMRG.  However the geometry in which DMRG is
typically run does not lend itself to the computation of entanglement spectra in which a multi-leg ladder system is divided into
subsystems, each consisting of several legs of the ladder.  One solution may be found in the approach taken in Refs.~\onlinecite{konik, james} where two dimensional systems are studied with a one dimensional DMRG algorithm with the caveat 
that the two dimensional systems is realized as an array of one dimensional continuum (not lattice) systems.  If we wish to
study fully two dimensional pure lattice systems, we will likely need recourse to more powerful but 
considerably more computationally demanding methods such as PEPS\cite{peps} or MERA.\cite{mera}  It has already been shown
that for certain types of translation invariant two dimensional models on an infinite lattice that the entanglement spectrum can be computed via the transfer matrix product operator.\cite{schuchladder}

In this manuscript we propose a tensor network method to describe quasi-one dimensional quantum systems with the help of tree tensor networks. The method is especially suitable to describe ladders with nonuniform coupling strengths while its computational advantage lies in the ability to optimize individual charge sectors efficiently for systems with a global symmetry. This allows us to achieve higher bond dimensions and thus higher accuracies. Most of all, it offers a convenient way to extract the entanglement spectra by a straight forward composition of the tensor network description for the reduced density operator for which the eigenvectors (and thus the entanglement spectrum) can be extracted using the existing tools such as the DMRG\cite{whiteprl92} or vNRG.\cite{vnrg}

Using this method we will revisit the connection between the entanglement spectrum and the energy spectrum of the real reduced system. We will study systems of two, three, and four leg ladders.  We will demonstrate that the entanglement spectra for three and four leg ladders is not simply related to the real spectrum of the subsystems resulting from a partition.  We will however confirm
the general thrust of Ref.~\onlinecite{albaprl} that the entanglement spectrum is affected by the boundaries separating the reduced system from the rest. 

\section{Method}
\begin{figure}
\centering
\includegraphics[width=0.9\columnwidth]{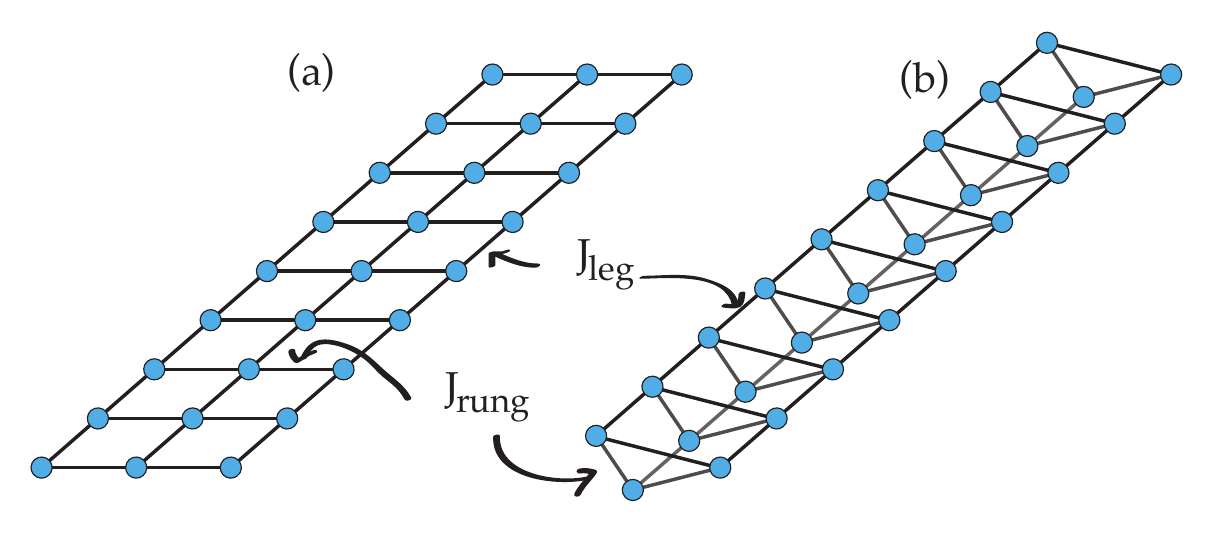}
\caption{Heisenberg model on a ladder with three legs with open (a) or periodic (b) boundary conditions on the rungs. }
\label{fig:ladder2}
\end{figure}

We consider a Heisenberg spin-$\frac{1}{2}$ model on a $m \times L$ ladder defined with Hamiltonian operator 
\begin{equation}
H = \sum_{i=1}^m \sum_{j=1}^L \Big( J_{\rm leg} \vec{\sigma}_{i,j} \cdot \vec{\sigma}_{i,j+1} + 
J_{\rm rung} \vec{\sigma}_{i,j} \cdot \vec{\sigma}_{i+1,j} \Big)
\end{equation}
where $\vec{\sigma} = (\sigma^x, \sigma^y, \sigma^{z} )$ denotes a vector of Pauli matrices.
We assume antiferromagnetic couplings and impose open boundary conditions on the legs (index $j$) and either open (Fig.~\ref{fig:ladder2}a) 
or periodic (Fig.~\ref{fig:ladder2}b) boundary conditions on the rungs (index $i$). 
We allow the couplings on the rungs and on the legs to be of a different  strength 
which we denote as $J_{\rm rung}$ and $J_{\rm leg}$, respectively. 
The model conserves the total $S^z$ quantum number given by 
$S^z = \frac{1}{2} \sum_{i,j} \sigma_{i,j}^z$ which allows us to consider different $S^z$ sectors 
separately. 
We shall focus only on the case $S^z = 0$ which is also the sector containing 
the (global)  ground state. 

The main objective of this work is to extract the entanglement spectrum, in particular the spectral gap, for a bipartite splitting of the ground state of the ladder into two parts along the longer axis, that is by cutting the rungs of the ladder. We shall put a special emphasis on the scaling of the entanglement spectral gap with the length of the ladder $L$ for various ladder widths $m=2,3,4$ and both open and boundary conditions on the rungs. Intuitively, one might expect that the spectral gap would be gapless for the $2\times L$ case, gapped for $4 \times L$ case and of yet to be studied nature for the $3\times L$, as a result of the Haldane conjecture.\cite{haldane} Specifically, a single Heisenberg spin-$1/2$ chain is gapless, so the entanglement spectrum for the $2\times L$ case should be gapless as well, since the $2\times L$ ladder can be split into two Heisenberg chains. A similar reasoning can be made for the $4\times L$ case.
In the case of a ladder with three legs, one would expect the entanglement spectrum to correspond to either of the two subsystems and would as such be gapped or gapless which would perhaps even depend on the ratio  $J_{\rm rung}/J_{\rm leg}$.
We will present numerical evidence, that this reasoning is not complete. Our results support a hypothesis that the entanglement spectrum depends not only on the subsystems but also on the boundary separating the subsystems.
While this argument plays no role for the $2\times L$ case, it makes a big difference for the 
$3\times L$ case depending on the boundary conditions on the rungs. 
In case of periodic boundary conditions, the subsystems are connected by two boundaries while only one boundary exists in the case of open boundaries on the rungs. 
Consequently, we have two different types of entanglement spectra for a $3\times L$ ladder, although the subsystems in a bipartite splitting are identical.
For the $4 \times L$ case the results suggest a gapped entanglement spectrum regardless of the boundary conditions on the rung.

The simulation of entanglement spectra requires first the ground state of the ladder, from which one could in principle extract the entanglement spectra directly using the singular value decomposition. In most cases, however, such an approach is exponentially hard and it is advantageous to obtain the reduced density operator by contracting over a subsystem and then simulate the eigenstates of the reduced density operator using some approximate method.

\subsection{Ground state simulation}
We shall simulate the ground state of the ladder with help of tensor networks by representing a quantum state of the system as a tree tensor network and then optimize a pair of sites or a single site at a time, such that the overall energy is minimized, and then proceed to optimize the next pair or the next site. The approach with optimizing two sites in a tree tensor network is new while the one-site approach has been used before.\cite{murg} However, in our framework we are able to fully exploit the symmetries which allows us to operate on the level of individual symmetry sectors and thus achieve a larger bond dimension in the network.

\begin{figure}
\centering\includegraphics[width=0.6\columnwidth]{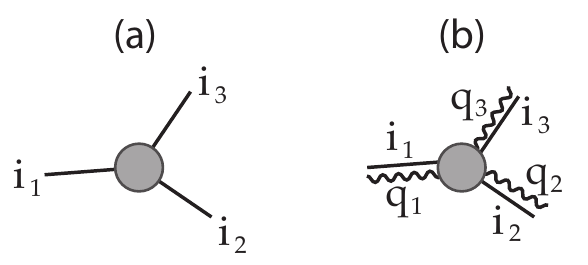}
\caption{A tensor corresponding to a node in a tree tensor network with at most three neighbors (a). With symmetries, the tensor is sparse (b).}
\label{fig:treetensor}
\end{figure}

Let us first describe a procedure to optimize any tree tensor network on an arbitrary geometry using variational principles. In order to reduce computational complexity we assume that each tensor in the network has at most three neighbors. Our scheme can easily be generalized to more than three neighbors, however, this would make the computational costs grow exponentially with the number of neighbors.  Furthermore, we make a crucial assumption that there a no loops in the tensor network. This requirement allows us to split the network into two subnetworks by cutting exactly one link. That said, we define a tree tensor network on nodes $\mu \in M$ by associating a tensor $A_{i_1,i_2,i_3}^{[\mu]}$ to the node $\mu$ (see Fig.~\ref{fig:treetensor} a).
The node $\mu$ is linked to three other nodes and the indices $i_1,i_2,i_3$ enumerate different virtual states on the links $1,2,3$, respectively. Not all nodes have three neighbors but some might have one or two open (unlinked) legs. To some of these nodes, let us call them $\mu^* \in M^* \subseteq M$, we associate the physical sites and one of the open links of the associated tensors plays the role of the physical index which is the local quantum number in the Hilbert space for the individual physical site. More specifically, a physical site $j$ is associated with the tensor $A^{[\mu(i)]}$ on the node $\mu(i)$ where the tensor leg $i_3$ is given by the configuration of the site, e..g $0$ for spin up and $1$ for spin down.  We shall use a formal notation 
$[\uA^{[\mu]}]_{i_1,i_2,i_3} = A^{[\mu]}_{i_1,i_2,i_3}$ when referring to tensors associated to nodes in the tree network and 
$[\uA^{[\mu^*] s}]_{i_1,i_2,i_3} = A^{[\mu^*]}_{i_1,i_2,s} \delta_{i_3,s}$ when explicitly referring to the tensors associated with the physical sites. 

Let us now write an ansatz for a quantum state $\Psi \in \mathcal{H}$ on an arbitrary lattice of $n$ sites by associating $n$ nodes of the network to the physical sites and adding some other nodes to connect the network. In total, we represent the quantum states with a tree network with $m \geq n$ nodes as 
\begin{equation}
\ket{\Psi} = \sum_{\underline{s}} {\rm Tr}\Big[ 
\prod_{\mu \in \mathbf{M} / \mathbf{M^*}} \uA^{[\mu]} 
\prod_{i} \uA^{[\mu^*(i)] s_i} 
 \Big] \ket{\underline{s}}
\label{eq:psi}
\end{equation}
where we have used an abbreviation $\underline{s} = (s_1,s_2,\ldots,s_n)$ and the
trace operation ${\rm Tr}[\bullet]$ should be understood as a tensor trace operation, i.e. summing over all indices on the links in the graph.

The loopless nature of the network makes it straightforward to include the symmetries in the ansatz of Eqn. (\ref{eq:psi}) by simply making the tensors $\uA^{[\mu]}$ sparse as shown in Fig.~\ref{fig:treetensor} b. To each link we associate an additional quantity, the charge, and require that the sum of all charges, flowing into the node in the network, equals to some constant value $Q = q_1 + q_2 + q_3$ which measures the total charge in the system. 
In the case of the spin model, the charges $(q_1,q_2,q_3)$ on the links $(1,2,3)$ would correspond to the total $S^z$ in the sub-graphs connected to the point $\mu$ by the links $(1,2,3)$, respectively. All charges should add up to the $S^z$ in the ground state. If the tensor $\mu$ corresponded to a physical node, then the leg $3$ would be open and $q_3$ would simply correspond to the local quantum number $S^z$ for the physical site. We note that in the contraction of the tensor network, we must contract a given charge $q$ with the conjugate charge, i.e. $q$ by $Q - q$.
Loopless nature of the tensor network also makes it possible to easily simulate fermionic systems without many modifications due to the fermionic signs. 
In such a case, the charges would correspond to the number of fermions flowing in from different parts of the network.
This representation of charge conservation differs from the usual representation in one dimensional system where the flow is conserved at each node, by introducing an outflowing charge which is a sum of all inflowing charges, including the local charge at the node, and imposing boundary conditions where no charge flows into the first site and the charge that flows out of the last site is equal to the total charge in the system. 
The representation we use is more convenient in the tree tensor network as it does not require to associate the direction of the flow (for a given node all charges flow into the node and they sum up to $Q$ at each node) nor specify the starting and the ending node, which makes is easier to consider completely generic tree networks without any regular topology. See also \cite{singh1,singh2} for a general treatment of symmetries in tensor networks algorithms.

An arbitrary linear map $G: \mathcal{H} \to \mathcal{H}$ can be represented as a sum of product linear maps $G = \sum_g O_g$ where $O_g = \prod_{j=1}^{n} o^{[g; j]}$ and each local operator $o^{[g; j]}$ acting on the physical site $j$ is linked to the physical leg of the tensor $\uA^{\mu^*(j)}$ as
\[
[o(\uA^{[\mu^*(j)]}) ]_{i_1,i_2,t} =  \sum_{s,t} o_{t,s}^{[g; j]} \uA^{[\mu^*(j)]}_{i_1,i_2,s}.
\]
This operation is completely local to the tensor $\mu^*(i)$ and as such does not increase the bond dimension of the tree tensor network. An expectation value $\bra{\Psi} G \ket{\Psi}$ can thus be obtained by summing up all contributions of the product operators $\sum_g \bra{\Psi} O_g \ket{\Psi}$  in parallel where, of course,  many contributions can be merged in the process, e.g the ones which act in the same sub-network and the like. 

The calculation of the expectation values therefore boils down to the calculation of scalar products $\braket{\tilde\Psi}{\Psi}$ and since the network is loopless, all the contractions can be done exactly without using any inverses. The approach is identical to the one for matrix product states and as such has been well studied in the literature. An important difference, though, is that the computational complexity of contracting the tree tensor network scales as $O(D^4)$ where $D$ is the maximum (typical) dimension of the bonds (i.e. number of indices on the links) as compared to $O(D^3)$ for matrix product states. In case we allowed four neighbors, the scaling would be $O(D^5)$ and so on.

In order to simulate the ground state, we must find a way to optimize individual tensor in the network, such that the total energy of the quantum state~(\ref{eq:psi}) is minimal. As in the case of one-dimensional systems, there are essentially two ways of doing that, the two-site (the DMRG way) and the one-site (the MPS way) optimization scheme, both of which have advantages and disadvantages.


\subsubsection{DMRG (two-site) optimization}
\begin{figure}
\centering
\includegraphics[width=0.9\columnwidth]{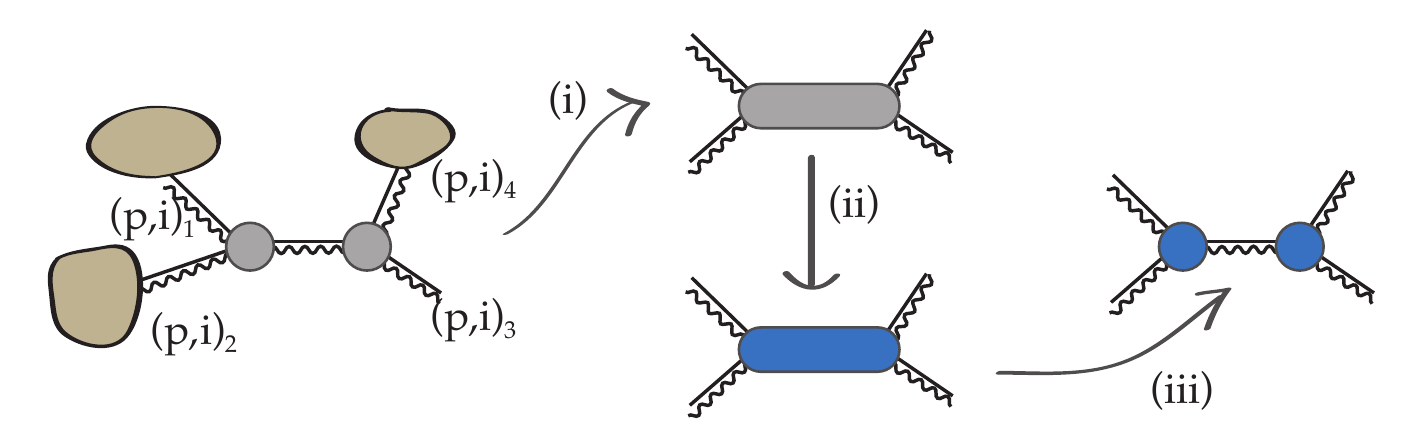}
\caption{Two site optimization scheme for tree tensor networks:
two neighboring nodes are merged to one block (i) which is replaced by the one which minimizes the total energy (ii), finally, the result is split back into two tensors (iii).}
\label{fig:pttn_dmrg}
\end{figure}

The two-site optimization scheme is best described on the sketch shown in Fig.~\ref{fig:pttn_dmrg}: we isolate two nodes in the network, merge the associated tensors into one bigger tensor (step i), optimize the bigger tensor such that the total energy of the system is minimized (step ii), and finally split the bigger tensor back to two smaller tensors while keeping the bond dimension under control (step iii). We repeat the procedure with the next pair of neighboring sites. This is the main ingredient of the well known DMRG algorithm to find ground states of quantum systems on a one-dimensional lattice. We will show that exactly the same principles can be used also with tree tensor networks, albeit with a higher computational cost. 

The advantage of the two-site optimization is that it generates the link between the two sites of interest from scratch and as such can create new charge sectors on the bonds or increase the number of kept auxiliary states (on the bond) if required. However, it requires a manipulation of a larger structure (a joint structure describing two nodes in the network) and, due to the nature of the Hamiltonian operator including hopping terms, it also requires handling several charge configurations simultaneously, albeit in a sparse way. Comparing the procedure in tree tensor networks (Fig.~\ref{fig:pttn_dmrg}) with the DMRG for one-dimensional lattices, we observe, that here the block of two sites can be connected to two, one, or none environments on either side. In the linear DMRG, the block is always connected to a single environment on either side. In addition, the environments are connected through the Hamiltonian, if it involves terms which operate on non-nearest neighbor terms in the tensor network.
When two neighboring physical sites are optimized, then either of the sites is connected to exactly one environment and the problem is translated exactly to the linear DMRG. 

Although the two site optimization has not been used before for optimizing tree tensor networks, the procedure is essentially the same as for the linear DMRG for which there already exists vast amount of literature.\cite{review,review1,review2}
Specifically to our case, we only mention that the joint tensor for two sites is optimized using a sparse Lanczos method where only 20 Lanczos steps are performed for each pair of sites, the reason being the computational cost. In the worst case, when the joint block of sites is connected to four environments (and the typical bond dimension being $D$), the computational cost of matrix-vector multiplication in the Lanczos algorithm scales as $O(D^5)$, compared to $O(D^3)$ for the linear DMRG algorithm. This makes the approach relatively expensive compared to the one-site optimization scheme described later. However, the advantage of creating new charge sectors and suppressing the insignificant ones makes the method very welcome for the initial stage of simulation, where a reasonable approximation for the ground state is obtained from a completely random initial realization of the tensors generating the quantum state. For such purposes, even $20$ Lanczos steps for each pair is sufficient to generate a good approximation for the ground state in two or three sweeps over the network.

\subsubsection{Variational (one-site) optimization}
The main optimization scheme we use to approximate the ground states of the ladders is the one-site optimization scheme. In this scheme, we choose a node in the network, optimize the associated tensor such that the total energy of the system is minimized, move to the next node and repeat the procedure. The advantage of this scheme is that we can reduce the optimization not just to a single tensor but to a single configuration block in the tensor, when the symmetries are used. This essentially means that we operate on the level of $D^3$ parameters whereas the total number of parameters describing the tensor $\uA^{[\mu]}$ for a given node $\mu$ is a factor of ten to hundred larger, depending on the allowed total bond dimension. A drawback of the scheme is, however, that we should already have a reasonable approximation for the ground state, otherwise we will spend unnecessary time (in the initial stage) optimizing charge sectors which in the end become completely irrelevant. 

\begin{figure}
\centering
\includegraphics[width=0.9\columnwidth]{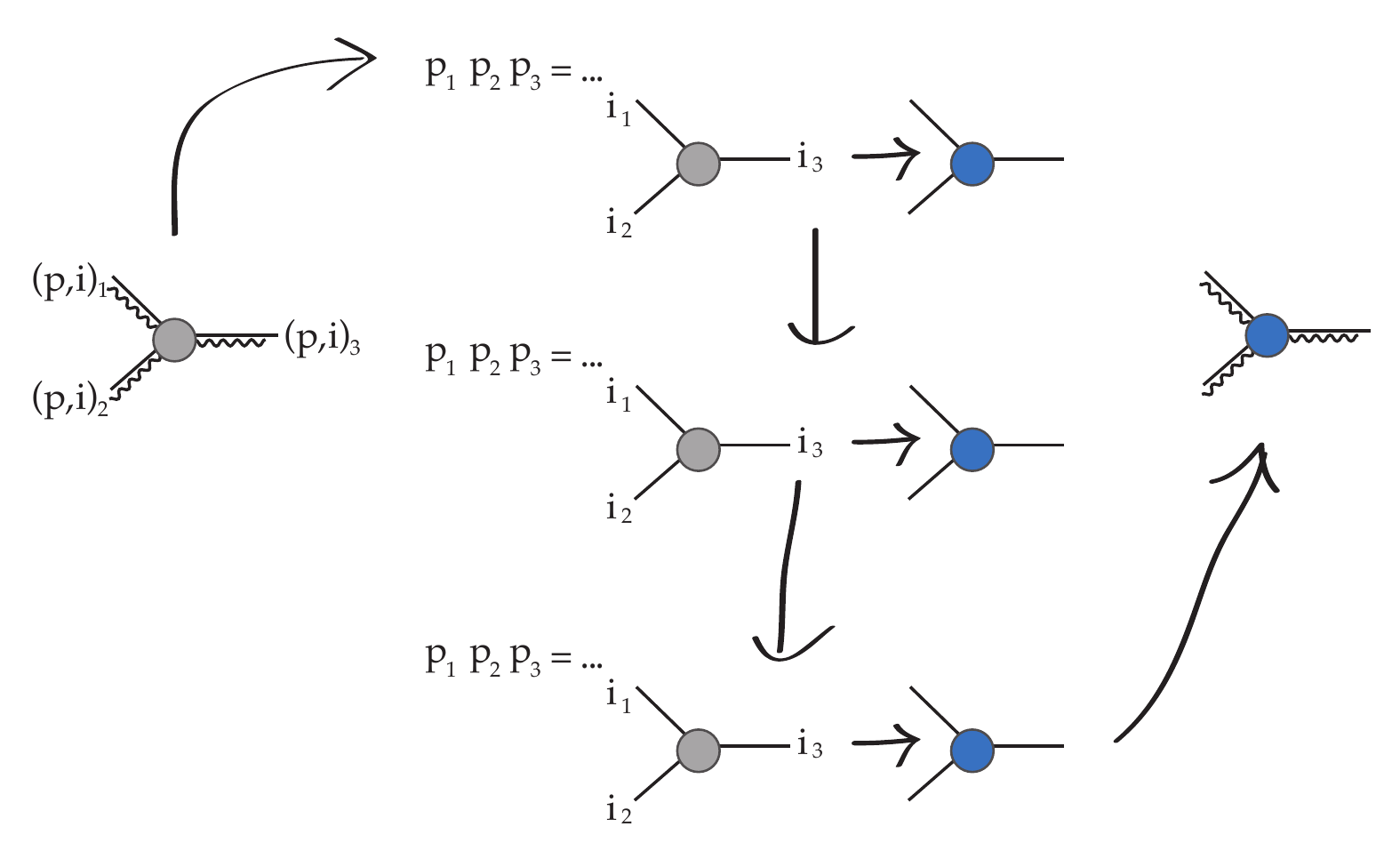}
\caption{One site optimization of a tree tensor network: one charge configuration of a single tensor in the tensor network is optimized at a time.}
\label{fig:pttn_vpopt}
\end{figure}

Let us now formally write the total energy of the system where we contract over all tensors in the network except for $\uA^{[\mu]}$. This leads to a description in terms of an ``effective'' hamitonian 
$H_{\rm eff}^{[\mu]}$ as
\begin{equation}
\bra{\Psi} H \ket{\Psi} = \sum_{(\underline{q},\underline{i})}
A_{(\underline{q'},\underline{i'})}^{*}
H^{[\mu]\, {\rm eff} }_{ (\underline{q'},\underline{i'}), (\underline{q},\underline{i}) }
A_{(\underline{q},\underline{i})}
\label{eq:Heff}
\end{equation}
Here we explicitly use the double index notation $\underline{q},\underline{i}$ which reflects the sparse nature of the tensors where $\underline{q} = (q_1,q_2,q_3)$ denote the charge configuration and $\underline{i}=(i_1,i_2,i_3)$ the dense tensor elements for this configuration $\underline{q}$.
We use the gauge transformations to transform the tensor network in such a form that the environment with respect to the node $\mu$ is unitary. This allows us to write the norm of the quantum state in a simple form 
\[
\braket{\Psi}{\Psi} =  \sum_{(\underline{q}, \underline{i} )}
A_{(\underline{q},\underline{i})}^{*}
A_{(\underline{q},\underline{i})}.
\]
Unless we are dealing with some trivial Hamiltonian which does not allow transfer of charges, then the charge configuration $\underline{p}$ at the node $\mu$ is coupled to some other charge configurations 
$\{ \underline{q} \}$ by the effective Hamiltonian 
$\underline{H}^{[\mu]\, {\rm eff} }_{ \underline{q}, \underline{q'} }$. 
We are therefore not allowed to optimize all charge configurations in parallel.
However, we can focus on one chosen configuration $\underline{p}$ and separate all terms 
in~(\ref{eq:Heff}) which contain $\underline{p}$ from those which do not.
For sake of brevity, let us represent the rank-3 tensors $\underline{A}^{[\mu]}_{\underline{p}}$ as 
vectors $a_{\underline{p}}$ (equivalent to stacking columns of a matrix to a long vector).
The result reads
\begin{equation}
\frac{\bra{\Psi}H\ket{\Psi} }{\braket{\Psi}{\Psi}}  = 
\frac{
a_{\underline{p}} \cdot \mathbf{H}^{\textrm{[eff]}}_{\underline{p},\underline{p}} a_{\underline{p}} + 
b_{\underline{p}} \cdot a_{\underline{p}}  + 
a_{\underline{p}} \cdot b_{\underline{p}}
+ f}
{a_{\underline{p}} \cdot a_{\underline{p}} + c^2}
\label{eq:totalE}
\end{equation}
where we have introduced the following quantities 
\begin{eqnarray}
b_{\underline{p}} &=&
\sum_{\underline{q} \neq p} \mathbf{H}^{[\textrm{eff}]}_{\underline{p},\underline{q}} a_{\underline{q}} \nonumber \\
f &=& \sum_{\underline{q}, \underline{r} \neq \underline{p} } 
b_{\underline{q}} \cdot \mathbf{H}^{[\textrm{eff}]}_{\underline{q},\underline{r}} a_{\underline{r}} 
\nonumber \\
c^2 &=& \sum_{\underline{q}\neq \underline{p}} a_{\underline{q}}\cdot a_{\underline{q}}.
\end{eqnarray}
This now allows us to optimize the total energy~(\ref{eq:totalE}) for individual charge configuration $\underline{p}$, after which we choose some other charge configuration $\underline{p'}$ until we explore all of them (see the sketch in Fig.~\ref{fig:pttn_vpopt}), at which point we reconstruct the whole tensor and move to the next node. The question however remains, how to optimize~(\ref{eq:totalE}) in an efficient way.

Let us assume that  $a_{\underline{q}} \in \mathbf{C}^{N}$. 
Obviously, $f \in \mathbf{R}$ and $c^2 \geq 0$. If $c^2=0$ then $a_{\underline{p}}$ is the only configuration for the tensor and hence $\vert\vert b_{\underline{p}} \vert\vert = f = 0$.
If $\vert\vert b_{\underline{p} } \vert\vert=0$, then the problem is transformed to an regular eigenvalue problem where the cost function (i.e. the energy) is minimized by the eigenvector of $\mathbf{H}^{[\rm eff}]$  with the smallest eigenvalue. We therefore assume that $\vert\vert b_{\underline{p}} \vert\vert > 0$ and in the following drop the charge notation $\underline{p}$.
Let us now consider an hermitian matrix $\mathbf{\tilde H} \in \mathbf{C}^{(N+1)\times (N+1)}$ with the
 matrix elements ${\tilde H}_{i,j} = H^{[\rm eff]}_{i,j}$ for $i,j=1,\ldots,N$, 
${\tilde H}_{N+1,j} = {\tilde H}_{j,N+1}^* = b_j$ and ${\tilde H}_{N+1,N+1} = f$. We also define a vector $y \in \mathbf{C}^{N+1}$ as $y_i = a_i$ for $i=1,\ldots,N$ and $y_{N+1} = 1$, and a diagonal matrix
${\tilde N}_{i,j} = \delta_{i,j}$ for $i=1,\ldots,N$ and ${\tilde N}_{N+1,N+1} = c^2$.
Obviously, the problem~(\ref{eq:totalE}) can now be written as an optimization problem
\[
\frac{\bra{\Psi}H\ket{\Psi} }{\braket{\Psi}{\Psi}}  = 
\frac{ y \cdot \mathbf{\tilde H} y }{y \cdot \mathbf{\tilde N} y }
\]
under the constraint that $y_{N+1} = 1$, known as the generalized eigenvalue problem. 
Note that we are not free to rescale the vector $a$ but we can always rescale the vector $y$. This problem is converted to a regular eigenvalue problem if $c^2 = 1$ when $\mathbf{\tilde N} = \mathbf{I}$. However, we can always (formally) rescale the whole tensor by $1/c$ in which case this would be true but now we also have to transform $f \to f/c^2$, $b \to b/c$ and $a \to a / c$. Finally, we solve the regular eigenvalue problem (by means of the exact diagonalization for small $N$ or the Lanczos algorithm for larger $N$), normalize the solution such that $y_{N+1} = 1$, and multiply the solution $a_i = y_i$ for $i=1,\ldots,N$ again with the factor $c$ to obtain the solution to the original problem~(\ref{eq:totalE}). 
In this procedure, we rely on a silent assumption that $c$ is not small. However, it $c$ was small, this would have a physical meaning that all the other charge configurations are negligible compared to the configuration $\underline{p}$ and the solution would negligibly differ from the solution if we set $b=f=c=0$.

To summarize the one-site optimization scheme, we optimize one charge configuration of a single tensor at a time and repeat the procedure iteratively for all charge sectors in the tensor and for all tensors in the tensor network, until the convergence is reached. The computational cost of each Lanczos step in the optimization (i.e. the cost of $\mathbf{H} x$) sums up to $(m + 4) 3 D^4$ with no additional prefactor where $m$ is the number of Hamiltonian terms which include the node $\mu$ (can be zero). This renders the optimization scheme very efficient in comparison to the two-site optimization scheme and it is thus possible to achieve significantly larger bond dimension $D$ than in the case when many charge sectors are combined into one large object.\cite{murg}

\subsection{Entanglement spectrum}

\begin{figure}
\centering
\includegraphics[width=0.9\columnwidth]{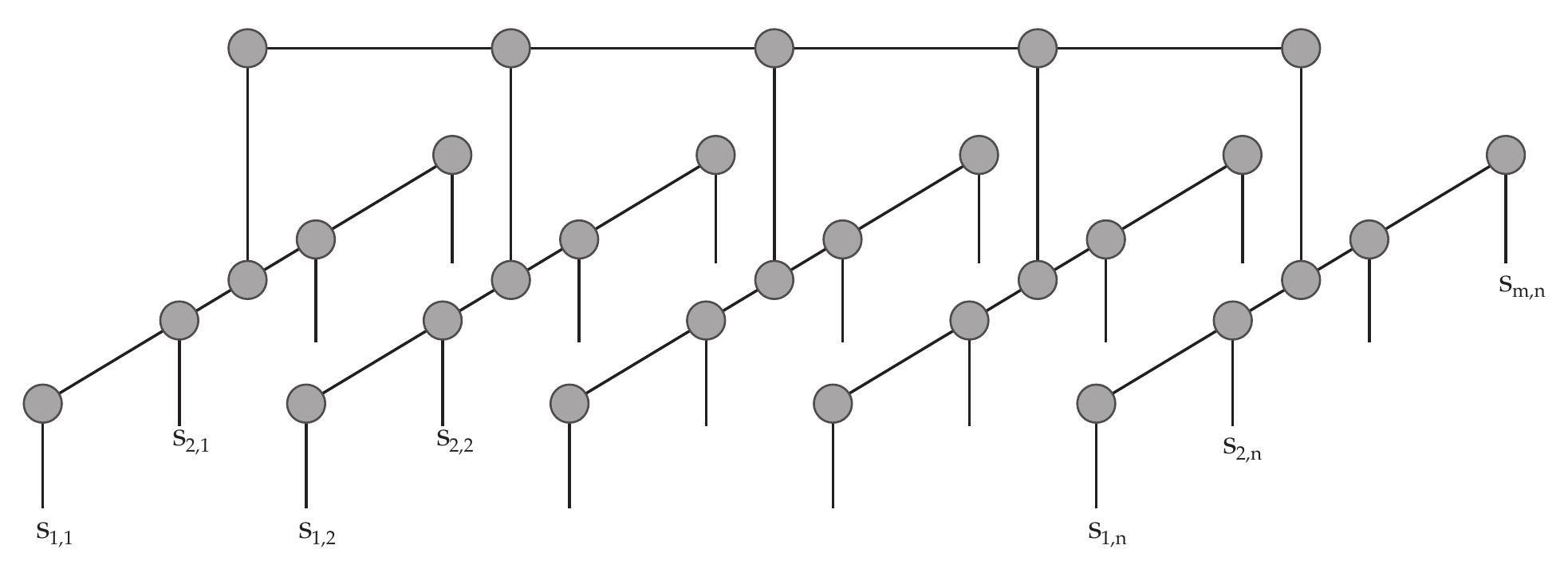}
\caption{Tree tensor network topology to represent quantum states on ladders as used in Ref.~\onlinecite{moukouri}.}
\label{fig:ladderttn}
\end{figure}

\begin{figure}
\centering
\includegraphics[width=0.9\columnwidth]{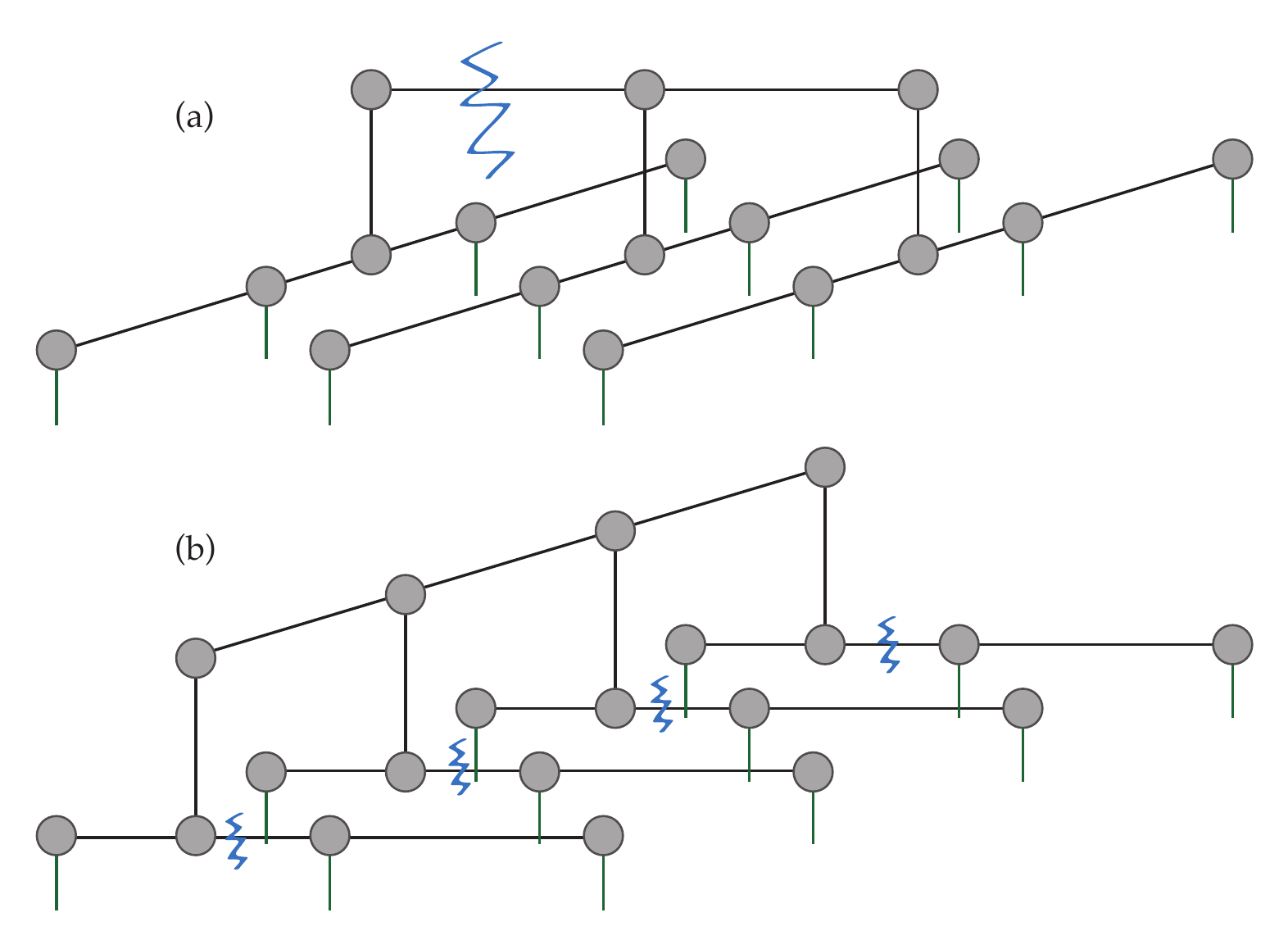}
\caption{Bipartite splitting of a ladder along the longer axis in a direct (a) and a rotated (b) geometry of the tree tensor network structure.}
\label{fig:bipartitesplit}
\end{figure}

Let us return to our original problem, that is to simulate the entanglement spectrum of $m \times L$ ladders for which we have to first calculate the ground state of the ladder. If the legs of the ladder are weakly coupled, then they act as effective systems which can be connected together in terms of a matrix product states. This is the formulation proposed in Ref.~\onlinecite{moukouri} where the ground states of $L \times m$ ladders (see Fig.~\ref{fig:ladderttn}) were simulated using a two-step DMRG scheme. In this scheme, the legs of ladder which are linear chains of length $L$ are considered as effective particles whose local basis is given by the excited states of the legs, computed by the DMRG with targeting several low energy excited states. In the next stage, the ground state of the ladder is obtained by simulating the ground state of a linear system of $m$ sites where each site is given by the effective description of the corresponding leg. While this approach offers a nice physical description in the case of scale separation, it lacks the feedback mechanism to refine the effective description of the legs and suffers from high computational complexity of the second simulation stage where the DMRG is performed on a system with a large local dimension. 
Nevertheless, the same geometry, depicted in Fig.~\ref{fig:ladderttn} can be treated as a loopless tree tensor network where the ground state can be simulated directly, without directly relying on the scale separation and, being an extension of matrix product states, offering a natural way for the feedback mechanism. 
Such a geometry of tensor network is physically well justified in the case considered in Ref.~\onlinecite{moukouri}, that is when the couplings between the legs (i.e. on the rungs) are fairly weaker than the couplings on the legs themselves, i.e. $J_{\rm rung} \ll J_{\rm leg}$. By using this topology, when two, three or four long legs are connected by a matrix product state (Fig.~\ref{fig:ladderttn}), it is trivial to obtain the entanglement spectrum by simply choosing a link on the top spin (see Fig.~\ref{fig:bipartitesplit} a), reorhorthogonalizing the network from both sides and performing a singular value decomposition on the resulting tensor. This however requires that the legs are only weakly coupled otherwise the required bond dimension on the top spin grows exponentially with the length of the ladder. 

We will be interested in the opposite scenario, when $J_{\rm rung} \geq J_{\rm leg}$, and in such a case the above described procedure is inefficient for $L > 10$ even for $2\times L$ ladders. In the end we will be interested in scaling of the spectral gap of the entanglement spectrum for which we shall require much larger systems for consideration. Since we shall only deal with ladders with up to $4$ legs, we shall exchange the role of legs and rungs and instead consider $L$ legs of length $m=2,3,4$. Now, however, obtaining the entanglement spectrum is far from trivial as cutting the system along the longer axis produces an exponentially large reduced density operator (Fig.~\ref{fig:bipartitesplit} b). Fortunately, as we shall see shortly, the density operator can also be considered as a matrix product operator for which we know how to extract the excited states, at least the largest ones, which in our case translates exactly to the largest eigenvalues of the entanglement spectrum. Another advantage of using the rotated geometry is that we can now easily consider periodic boundary conditions on the legs because there are only at most physical $4$ sites on each leg and the increase on the bond dimension due to coupling of the two boundary sites, is negligible, if any.
In order to avoid confusion, we shall continue using the nomenclature defined in the introduction where the long chains are called legs and the connections between them are called rungs as shown in Fig.~\ref{fig:ladder2}; the rotation of the ladder should only be regarded as a technical trick.

\begin{figure}
\centering
\includegraphics[width=0.9\columnwidth]{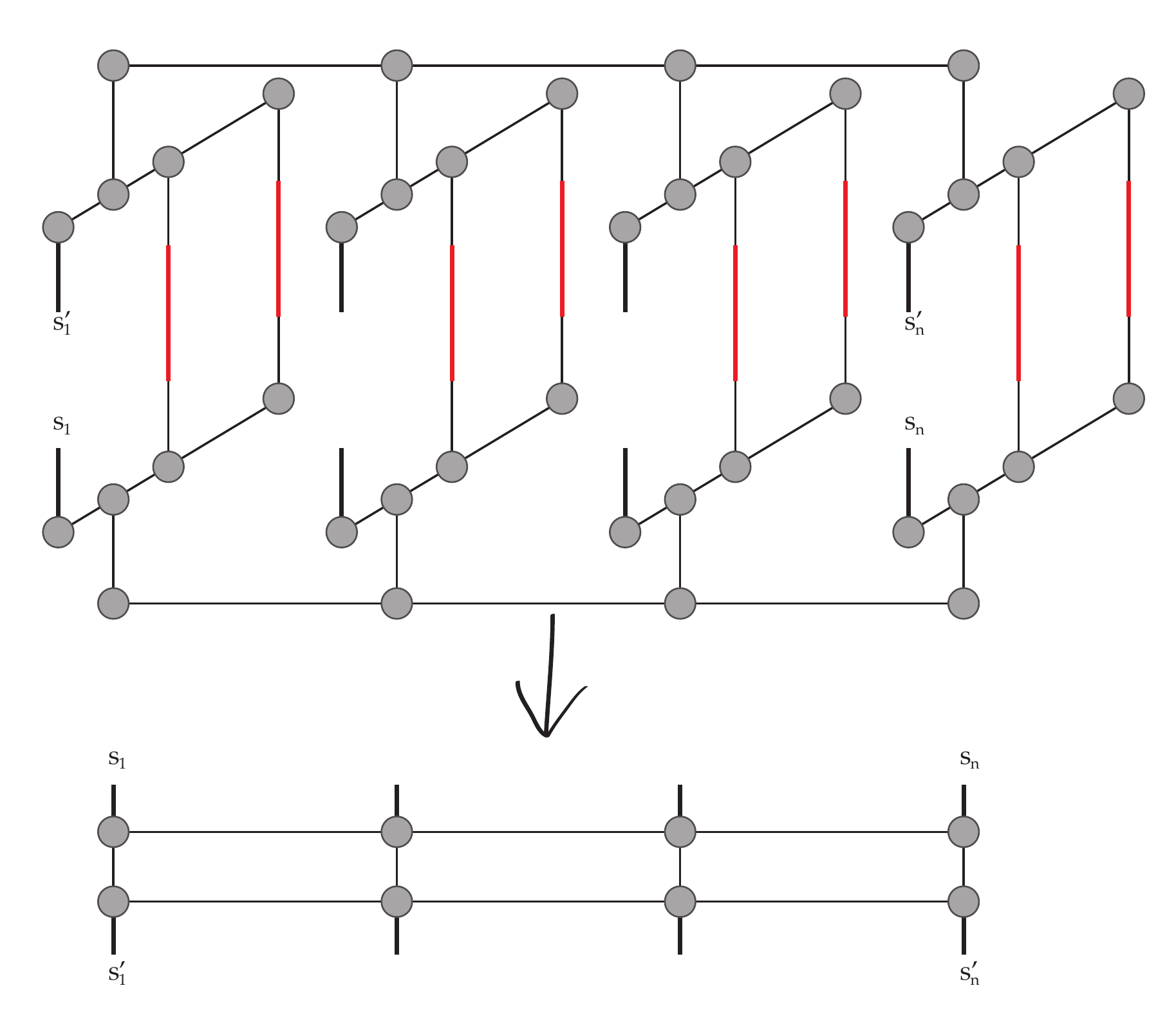}
\caption{The reduced density operator is obtained by tracing over physical degrees of freedom of the environment.}
\label{fig:rhompo}
\end{figure}

Let us now assume we have obtained the ground state of the ladder in a form depicted in Fig.~\ref{fig:bipartitesplit} b. We can formally decompose the ground state into two parts where the part left of the cut in Fig.~\ref{fig:bipartitesplit} b is called ``the system'' and the part right of the cut is called ``the environment'', by the so called Schmidt decomposition
\[
\ket{\Psi} = \sum_k \ket{\psi^{[S]}_k } \ket{\psi^{[E]}_k}
\]
where $\{ \psi^{[S]}_k \}$ and $\{ \psi^{[E]}_k \}$ are orthogonal sets in the subsystems $S$ and $E$.
The reduced density operator is obtained by tracing the full density operator $\ket{\Psi}\bra{\Psi}$ over the environment, spanned by an orthonormal set $\{ \phi^{[E]}_k \}$, 
\[
\rho_S = \sum_{j} \bra{\phi^{[E]}_j} \Big( \sum_{k,l} \ket{\psi^{[S]}_k } \ket{\psi^{[E]}_k} \bra{\psi^{[S]}_l } \bra{\psi^{[E]}_l}  \Big) \ket{\phi^{[E]}_j} 
\]
which can be further simplified using the fact that $\sum_j \ket{\phi^{[E]}_j} \bra{\phi^{[E]}_j}   = \mathbf{1}^{[E]}$ to 
\begin{equation}
\rho_S = \sum_{k,l} \braket{\psi^{[E]}_l}{\psi^{[E]}_k}  \ket{\psi^{[S]}_k } \bra{\psi^{[S]}_l }.
\label{eq:rho}
\end{equation}
The reduced density operator is thus simply obtained from the ground state by contracting over the physical degrees of freedom in the environment which is schematically shown in Fig.~\ref{fig:rhompo}. 
The result of the contraction is a product of two matrix product operators or, finally, 
a single matrix product operator. The quantum state describing the ground state was normalized and thus no additional normalization factor appears in~(\ref{eq:rho}). Therefore, the eigenvectors and eigenvalues of the reduced density operator are precisely the eigenstates of the matrix product operator depicted in Fig.~\ref{fig:rhompo}. This operator, however, is obtained from a double layer tree tensor network structure and the bond dimension in its MPO (matrix product operator)
representation can be very large. In order to simulate its eigenvectors we have to first truncate it to a manageable size by first eliminating redundant auxiliary degrees of freedom (by means of matrix factorization) and then truncating the auxiliary state by retaining at most $1000$ states on each bond (by means of a singular value decomposition). 
This truncation procedure is a standard ingredient in all MPS simulations (see e.g. Ref.~\onlinecite{verstraetempo}). The reduced density operator retains the symmetries of the quantum state and as such preserves the total number of particles or, in our case, the total $S^z$ of the subsystem. The tensors in the MPO are therefore sparse. 

Having obtained the reduced density operator in a form of a MPO, we can calculate the corresponding eigenvalues and eigenvectors. One way to accomplish this task is to use the DMRG algorithm \cite{whiteprl92} which can, despite not frequently used for that purpose, target not just the ground state but several excited states, with the only condition that the operator is hermitian. This is indeed the case for the density operator and we can compute a few of its eigenvalues and eigenvectors by simply plugging $H_{\rho} = - \rho$ to a ready-to-use implementation of the DMRG algorithm with targeting (with slight modifications for the support of matrix product operators).
The DMRG with targeting proves to be fairly expensive, especially when the local dimension is larger than $2$ and when the Hamiltonian itself is fairly complicated. In comparison to calculating the excited states of a Heisenberg chain of length $L$, calculating the same number of eigenstates of the reduced density operator for  a $2\times L$ ladder is a hundred to thousand times more expensive, simply because we are dealing with a nonlocal MPO with a bond dimension $1000$ instead of a local  MPO with a bond dimension $5$ (i.e. for the Heisenberg model); an additional factor stems from the fact that it is easy to separate the charge sectors in the Heisenberg model whereas in our case, the MPO (after truncation) has a more complicated structure and we have to optimize larger blocks at a time. 

An alternative way to simulate the excited states is to use the variational NRG method \cite{vnrg} which can be seen as a one-site version of the DMRG with targeting. It can be used in place of the DMRG or as an additional stage in the simulation to optimize the results obtained by the preceding DMRG simulation. In both cases, the excited states can be described by the following Ansatz
\begin{eqnarray}
\ket{\psi_k} &=& \sum_{s_1,\ldots,s_n} {\rm tr}[\mathbf{L}^{[1] s_1}  \cdots \mathbf{L}^{[n/2] s_{n/2}} 
\mathbf{X}^{k} \mathbf{R}^{[n/2+1] s_{n/2+1}}  \mathbf{R}^{[n] s_{n}}  ] \nonumber \\
&\times& \ket{s_1,s_2,\ldots,s_n}
\label{eq:cpnrg}
\end{eqnarray}
with unitary constraints 
$\sum_{i,s} L_{i,j}^s L_{i',j'}^{s} = \delta_{j,j'}$, 
$\sum_{j,s} R_{i,j}^s R_{i',j'}^{s} = \delta_{i,i'}$ and 
$\sum_{i,j} X_{i,j}^k X_{i,j}^{k'} = \delta_{k,k'}$. 
These constraints guarantee that the states in ~(\ref{eq:cpnrg}) form an orthonormal set. 
The matrices $\mathbf{L}^{[j] s_j}$, $\mathbf{R}^{[j] s_j}$ and $\mathbf{X}^{k}$ are
then optimized by minimizing the cost function
\begin{equation}
f(\mathbf{L}, \mathbf{X}, \mathbf{R}) = - \sum_k \bra{\psi_k} \rho \ket{\psi_k} = \textrm{min}.
\label{eq:cost}
\end{equation}
The method was described in detail in Ref.~\onlinecite{vnrg}.
We start with some initial realization of matrices $L$, $X$ and $R$
and update them site by site by minimizing the cost function~(\ref{eq:cost})  
under unitary constraints. 
In this way the set of states described by the ansatz of Eqn.(\ref{eq:cpnrg}) remains orthonormal at all times.
Eventually we end up with an approximation to the eigenstates with 
the smallest eigenvalues which in our setting translates to the singular vectors corresponding to the 
largest singular values. 
The complexity of this simulation scales as $O(D^3)$ where $D$ is the 
maximal bond dimension in the Ansatz~(\ref{eq:cpnrg}); the optimization of the central tensor 
$\mathbf{X}^{k}$ scales as $O(D^3 m)$ where $m$ is the number of the excited states $\psi_k$ 
described by~(\ref{eq:cpnrg}). 
For our purposes we choose $m=10$ which gives us ten largest Schmidt coefficients and their singular vectors. This in turn gives us the entanglement spectrum.
Similarly to the ground state simulation, it is advantageous to initialize the tensors in~(\ref{eq:cpnrg}) by performing one sweep of two-site DMRG simulation with targeting which eliminates insignificant symmetry sectors and thus reduces the computational costs of the one site optimization in the next stage.

In principle, if we only wanted to calculate the spectral gap between the sectors $S^z=0$ and $S^z=1$ and not higher excited states in the entanglement spectrum, it would suffice to calculate the ground state of the matrix product operator (e.g. using standard DMRG) for the reduced density operator in both subsectors.

\section{Results}
We shall use the methods described in this manuscript to calculate the low-lying entanglement spectrum of $m \times L$ ladders for $m=2,3,4$ with either open or periodic boundary conditions on the rungs (Fig.~\ref{fig:ladder2}). The simulation proceeds in three steps:  we calculate the ground state of the ladders from which we form a matrix product operator describing the reduced density operator in a bipartite splitting along the long axis. Finally, to obtain the low lying entanglement spectra,
we simulate a few  eigenvectors of this matrix product operator corresponding to the largest singular values.

\subsection{Ground state of the ladders}
\begin{figure}
\centering
\includegraphics[width=0.9\columnwidth]{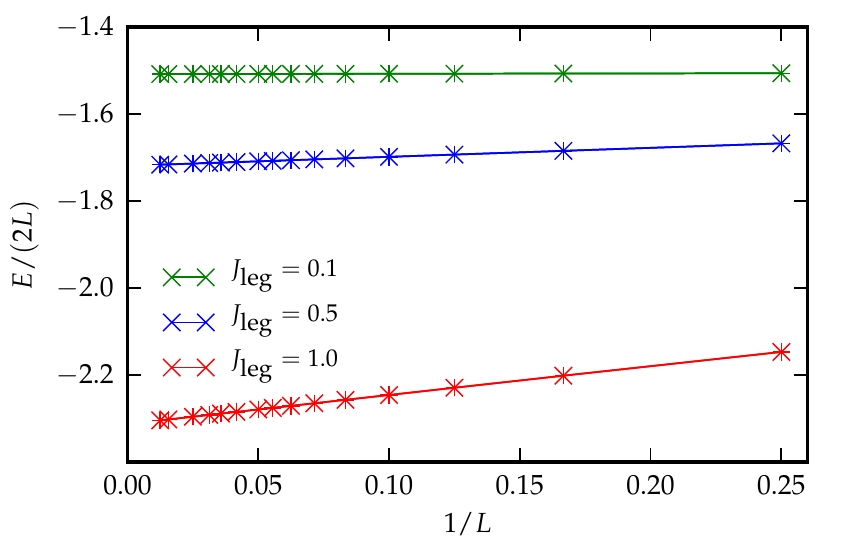}\\
\includegraphics[width=0.9\columnwidth]{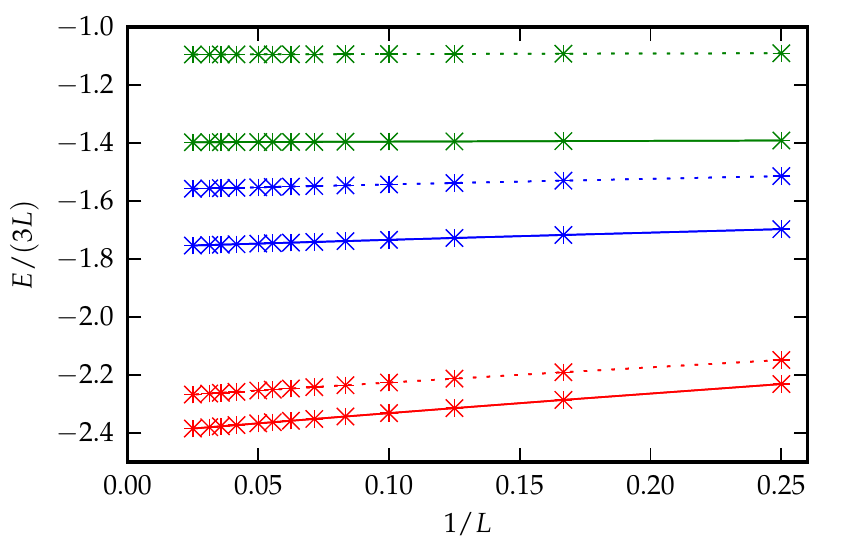}\\
\includegraphics[width=0.9\columnwidth]{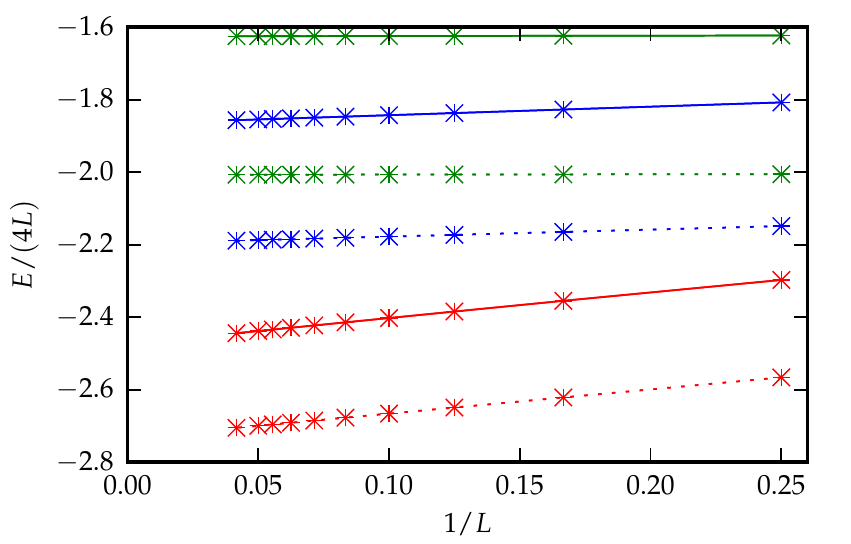}
\caption{Ground state energy per site for $m\times L$ ladders with $m=2,3,4$ (top to bottom) with rung-periodic (dotted lines) and open (solid lines) boundary conditions. In all cases $J_{\rm rung}=1$.
Bond dimensions $D$ used:  20 (cross), 30 (plus).}
\label{fig:plotgs}
\end{figure}
The ground state of the ladders is obtained in two steps. First we find a crude approximation for the ground state by using the two-site optimization scheme (DMRG-like) and then optimize the results using the variational one-site optimization scheme. In all cases, we use an ansatz with symmetries restricting to $S^z=0$ subsector.  As mentioned previously, we in fact described the tree tensor network in a rotated geometry (Fig.~\ref{fig:bipartitesplit} b), since we consider the ratios between the couplings $J_{\rm rung}/J_{\rm leg} = 1, 2, 10$. 

In the simulation we keep at most $1000$ overall states for each bond in the tree tensor network and at most $D$ states in each charge sector where $D=20,30$.  The results are presented in Fig.~\ref{fig:plotgs} where the points connected by solid lines correspond to open boundary conditions and the dashed lines to the periodic boundary conditions on the rungs.
The bond dimension $D$ used in simulations is denoted by various symbols: cross for $D=20$ and plus for $D=30$. In all cases, the symbols essentially overlap and no visible difference can be observed. 
\begin{figure}
\begin{center}
\begin{tabular}{|l|l|c|c|c|}
\hline 
$J_{\rm leg}$ & b & $2 \times L$ & $3 \times L$ & $4 \times L$  \\
\hline
$0.1$ & o & $-1.50786650301$  & $-1.39739531959$  & $-1.62535055215$ \\
$0.5$ & o & $-1.71869127504$ & $-1.75880276836$ & $-1.86629861886$\\
$1.0$ & o & $-2.31201208933$ & $-2.40018707942$  & $-2.473345887054$ \\
\hline
$0.1$ & p & & $-1.09465154452$ & $-2.00733345726$ \\
$0.5$ & p & & $-1.56190824628$ & $-2.19712045516$ \\
$1.0$ & p& & $-2.27972332710$ & $-2.73289318696$ \\
\hline
\end{tabular}
\end{center}
\caption{Extrapolated normalized ground state energies $E_0/ (m \times L)$ for $L\to\infty$ with 
$m=2,3,4$ and open (o) or periodic (p) boundary conditions. We used $D=30$ as the bond dimension.}
\label{fig:extrapolation}
\end{figure}
In Figure~\ref{fig:extrapolation} we give a table of normalized energies for a fixed bond dimension $D$ obtained by extrapolation the data in the Fig.~\ref{fig:plotgs} to $L\to\infty$. The results agree with Ref.~\onlinecite{riera} where the extrapolated energy for a $2 \times L$ ladder with $J_{\rm rung}/J_{\rm leg}=1$ was found to be $\lim_{L\to\infty} E_0/(2 L) = -2.312$ (in our units).

We shall briefly mention the computational parameters used to obtain the results in Fig.~\ref{fig:plotgs}. 
The initial two-site (DMRG-like) optimization was done by sweeping twice over the network where at each step the joint tensor was optimized by performing at most $30$ Lanczos iterations. 
The result of the DMRG simulation was used as an initial state for the variational one-site optimization which was performed until the relative difference between the energies after two consecutive \emph{sweeps} became less than $10^{-14}$. The ground state for smaller systems (all $2\times L$ and up to $3\times 20$ and $4\times 12$) can be obtained in less computational time by forgoing the two-site scheme and starting with the one-site scheme on a random realization of $\Psi$. However, for larger systems, it is advantageous to initialize the state by the DMRG which eliminates the unneeded charge configurations. The computational time required to obtain the results shown in Fig.~\ref{fig:plotgs} is in the range of a few seconds to 3 hours in the worst case.

\subsection{Entanglement spectrum}
\begin{figure}
\centering
\includegraphics[width=0.9\columnwidth]{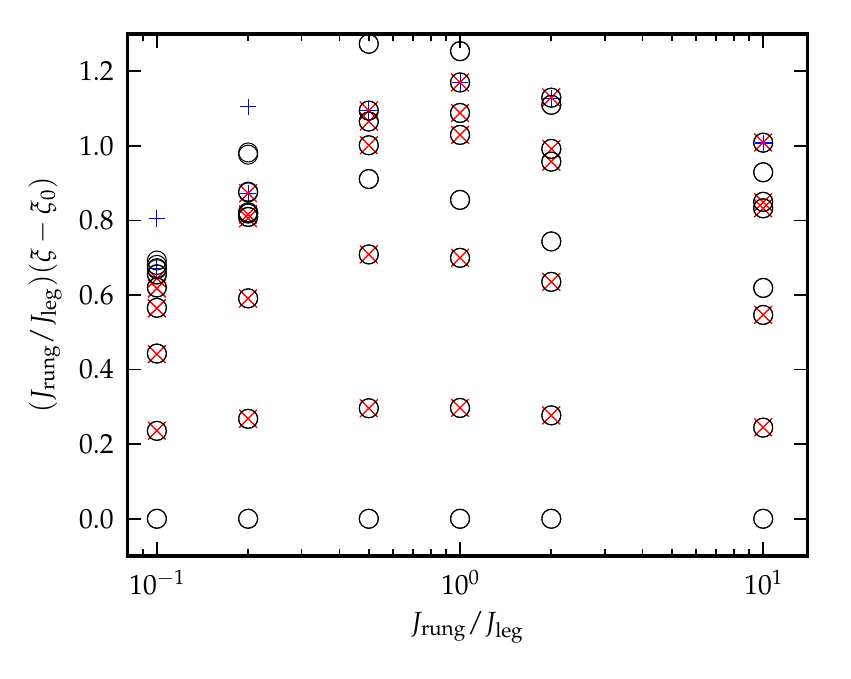}
\caption{The entanglement spectrum for a $2 \times 32$ ladder as a function of $J_{\rm rung}/J_{\rm leg}$ and various $S^z$ sectors of the subsystem: $S^z=0$ (black circle), $S^z=1$ (red cross) and $S^z=2$ (blue plus).
}
\label{fig:plotesp32}
\end{figure}

As a benchmark of the method we calculate the entanglement spectrum in a bipartite splitting of a $2\times L$ ladder for various ratios $J_{\rm rung}/J_{\rm leg}$. The reduced density operator is invariant with respect to $S^z$ in the subsystem and we simulate the most significant values separately for $S^z=0,1,2$. The results shown in Fig.~\ref{fig:plotesp32} are essentially a reproduction of the results presented in Ref.~(\onlinecite{poilblanc}) but 
for open boundary conditions on the legs and a slightly larger system size. The entanglement spectra shown in Fig.~\ref{fig:plotesp32} however differ from Ref.~\onlinecite{lauchli} due to different boundary conditions on the legs. Confirmed by an exact diagonalization for a $2 \times 10$ ladder, we find that the second excited state is a triplet for open boundary conditions as opposed to a singlet for periodic boundary conditions on the legs.

The entanglement spectrum $\{ \xi_j \}$ was obtained from the eigenvalues of the reduced density operator $\{ \rho_j \}$ as $\xi_j = -\log \rho_j$. For presentation purposes, the ``ground state'' (i.e. the lowest state for $S^z=0$) was subtracted from the spectra and the result was multiplied by the ratio $J_{\rm rung}/J_{\rm leg}$. The charge sectors ($S^z$) are  given by the shapes of the symbols: a circle for $S^z=0$, a cross for for $S^z=1$ and a plus for $S^z=2$.  Each point is replotted several times corresponding to the different bond dimensions of the underlying ground state ($D$) and maximal number of states in the representation of the MPO. Ideally the points should overlap.  When they don't, we obtain an
idea of the uncertainty in the determination of the spectra.
The structure of the low lying entanglement spectrum (singlet, triplet, \ldots) agrees with the structure of the energy spectrum for a one-dimensional Heisenberg spin-$1/2$ model.
We observe that the lowest levels of the entanglement spectra are well represented for any simulation parameters whereas the excited states require more computational power and are less precise, the fact known already from the method of DMRG with targeting many excited states.

\subsection{Entanglement spectral gap}
\begin{figure}
\includegraphics[width=0.9\columnwidth]{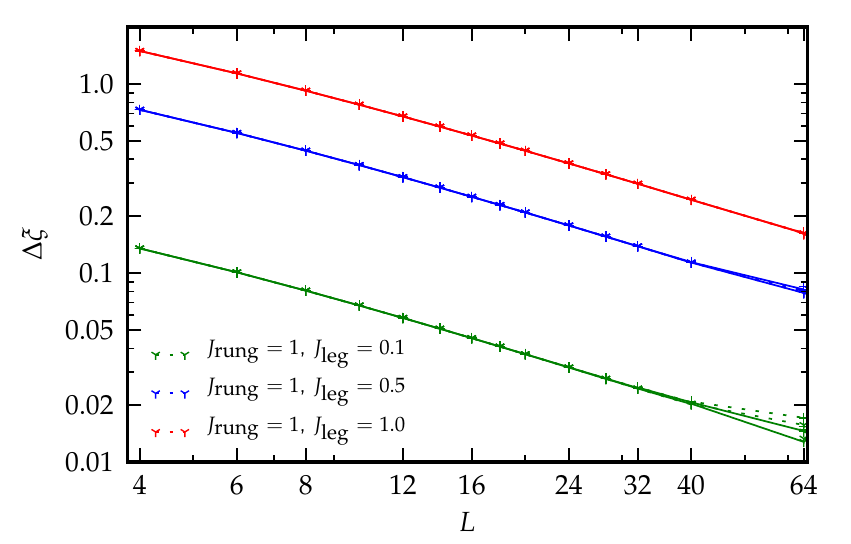}
\caption{Entanglement spectral gap in a bipartite splitting of $2\times L$ ladders into two chains of length $L$.}
\label{fig:gap_2xL}
\end{figure}

We will focus in the remainder of this section on computing the entanglement gap.  We will be interested in particular in investigating
whether the presence of an entanglement gap implies a presence of a gap in the actual spectrum of the subsystem arising from the partition
(and vice versa).  
We know that this
is the case of two-leg ladders where it was shown in Ref. \onlinecite{poilblanc} that the entanglement spectra associated with
the dividing the ladder into two chains mimicked that of the actual spectrum of an individual chain.  
We verify this observation by our simulations where we observe 
(Fig.~\ref{fig:gap_2xL}) that the spectral gap vanishes for $N\to\infty$ for all considered 
ratios $J_{\rm rung}/J_{\rm leg}$. Here, the symbols denote the  bond dimension of the singular vectors~(\ref{eq:cpnrg}) whereas the connecting line denotes the bond dimension of the corresponding ground state (dashed for $D_{\rm vnrg}=10$, solid for $D_{\rm vnrg}=20$, dotted for $D_{\rm vnrg}=30$). 
The results practically overlap in all cases.

In all cases, including those that follow, we set the bond dimension for the singular vectors (i.e. the eigenvectors of the reduced density operator) described by the ansatz in Eqn. (\ref{eq:cpnrg}) to $D_{\rm vnrg}=10,20,$ and $30$ and no limit was imposed on the total bond dimensions (sum over all charge sectors). 
We considered three different maximal bond dimensions for the matrix product operator representation of the reduced density operator:
$700$, $900$, and $1100$.  The corresponding ratio between the minimal allowed singular value and the maximal one in the bipartite splitting of the MPO was $10^{-7}$, $10^{-9}$, and $10^{-11}$, respectively. 

\begin{figure}
\centering
\includegraphics[width=0.9\columnwidth]{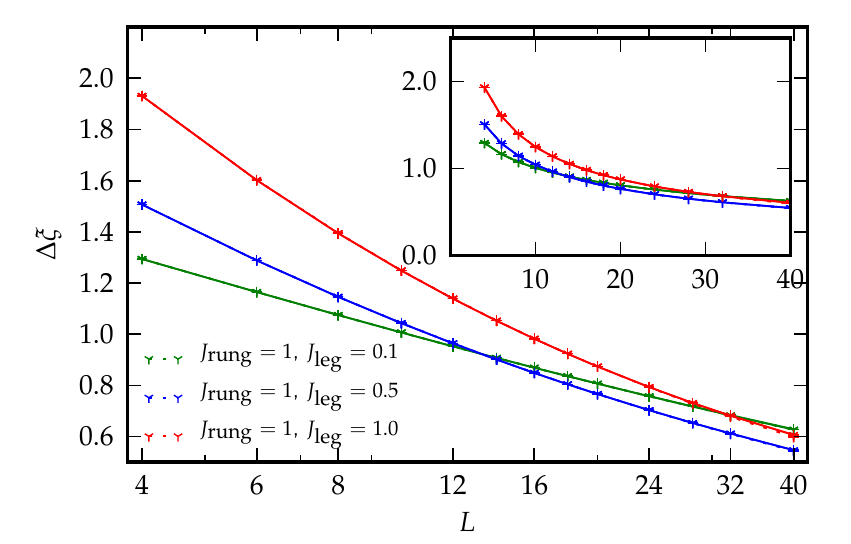}
\includegraphics[width=0.9\columnwidth]{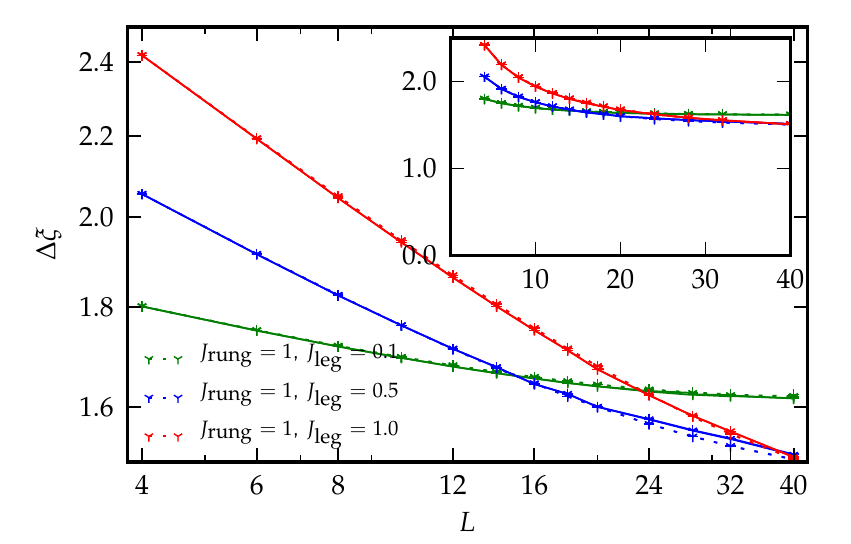}
\caption{Entanglement spectral gap for the $3 \times L$ ladder with open (top) and periodic (bottom) boundary conditions on the rungs; the main plots are in the log-log scale, the insets in the normal scale.}
\label{fig:gap_3xL}
\end{figure}
\begin{figure}
\centering
\includegraphics[width=0.9\columnwidth]{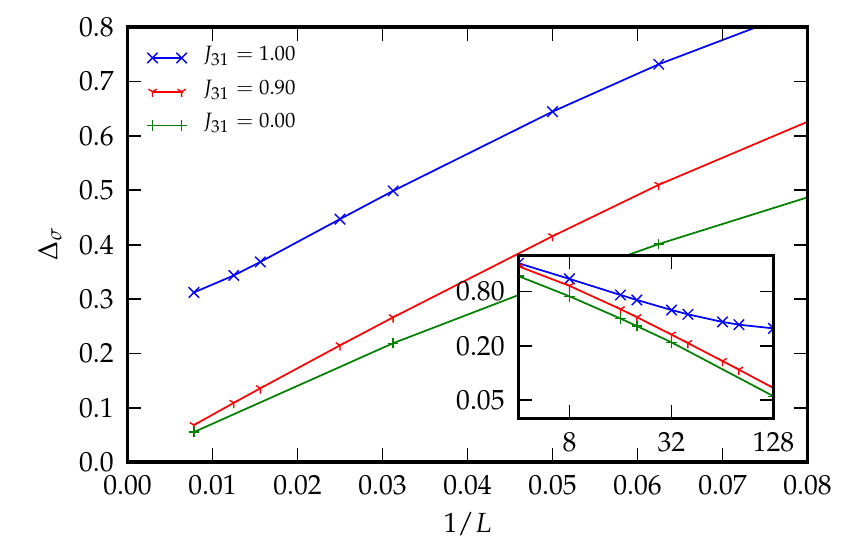}
\caption{Spectral gap of the $3\times L$ ladder with asymmetric boundary condition on the rungs, $J_{12}=J_{23}=1$ and various $J_{31}$. In all cases, $J_{\rm leg}=0.5$. The inset shows the spectral gap versus the system size in a log-log scale.}
\label{fig:3xLx}
\end{figure}
In the case of $3\times L$ ladders it is not clear what to expect as we divide the ladder into a chain and a two-leg ladder.  A two-leg Heisenberg 
spin ladder (with anti-ferromagnetic interactions) is expected to effectively behave as a Heisenberg spin chain of an integer spin and thus display a gap in the thermodynamic limit whereas the gap of a single Heisenberg spin-$1/2$ vanishes in the thermodynamic limit.  It is not {\it a priori} clear then which of these two
options the entanglement gap will mimic.  Surprisingly, the results depend on the particular boundary conditions we impose on the
rung.   We see this 
in the numerical results shown in Fig.~\ref{fig:gap_3xL} where we observe a clear difference between open boundary conditions and periodic
boundary conditions. In the case of open boundary conditions (top) where it suffices to make a single cut to separate the three leg ladder into two parts, we observe a similar behavior as in the $2 \times L$ case, that is a vanishing gap for $L \to \infty$. However, in the case of periodic boundary conditions on the rungs (bottom) the gap remains finite for $L\to\infty$. 

This result is less surprising once we consider that it has been shown\cite{nishimoto,sakai} that the spectrum of the
$3 \times L$ ladder itself depends on the boundary conditions.  With periodic boundary conditions, frustration is present on
each of the rungs on the ladder and the system is seen to be gapped.  As soon as the frustration is removed by making one bond on the rung weaker or stronger, the system becomes gapless.
We confirm this behavior by computing the gap of the $3 \times L$ ladders were $J_{\rm rung}$ are chosen as $(1,1,J_{31})$ with $J_{31} =0, 0.9, 1$ and $J_{\rm leg} = 0.5$ .   The results shown in Fig.~\ref{fig:3xLx} confirm that the system is gapped at $J_{31}=1$ and gapless otherwise.
Thus we see for the three-leg ladders the entanglement gap does not necessarily mimic that of the subsystem arising from the partition, but
rather follows the full system itself.

The behavior of the entanglement for three-leg ladders has implications for the perturbative (in $J_{\rm leg}$) entanglement Hamiltonian.
At zeroth order in $J_{\rm leg}$ the entanglement Hamiltonian for a single chain (supposing we trace out two legs of the three-leg ladder)
must be equal to the identity,
\begin{equation}
H_{\rm entanglement}  = a_0{\bf I} + {\cal O}(J_{\rm leg}),
\end{equation}
where $a_0$ is some constant
This follows by SU(2) invariance and that the only SU(2) invariant operator involving operators sitting at a single site is the
identity.  At next order, SU(2) invariance gives the entanglement Hamiltonian for the chain as
\begin{equation}
H_{\rm entanglement}  = (a_0+J_{\rm leg}b_0){\bf I} + b_1 J_{\rm leg}\sum_{i} {\bf S}_i\cdot{\bf S}_{i+1} + {\cal O}(J^2_{\rm leg}),
\end{equation}
i.e. the Heisenberg Hamiltonian and where $b_{0,1}$ are constants.  
This follows as the first order entanglement Hamiltonian must involve terms which
are no more non-local than nearest neighbor.  However such a Hamiltonian is necessarily gapless.  Thus in
order to produced a gapped entanglement Hamiltonian (as we find for the case of PBCs) we must consider the $J^2_{\rm leg}$
contribution to it.  While it is beyond our ability to easily compute this correction, it will involve next nearest interaction terms that for the
case of PBCs lead to a gapping out of the spectrum.

\begin{figure}
\centering
\includegraphics[width=0.9\columnwidth]{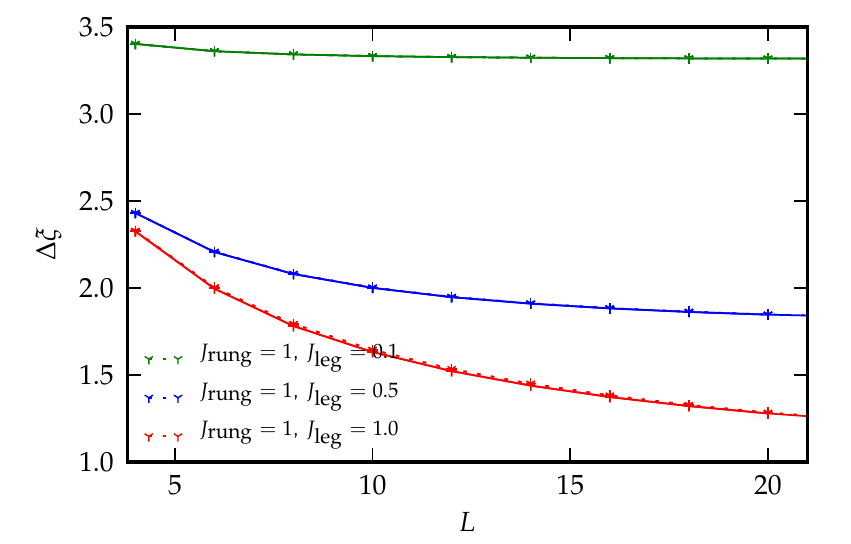}
\includegraphics[width=0.9\columnwidth]{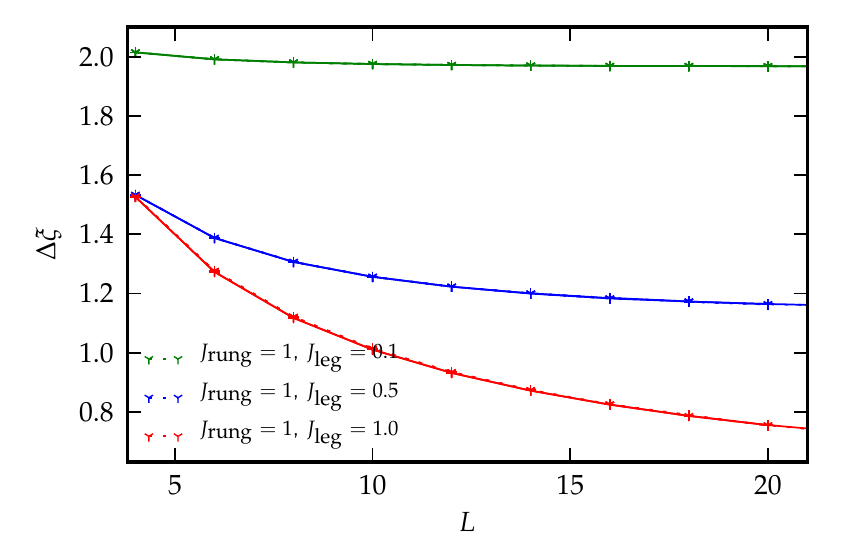}
\caption{Entanglement spectral gap for a bipartite splitting of a $4\times L$ ladder into two $2\times L$ ladders for the ground state of the ladder Heisenberg Hamiltonian with open (top) and periodic (bottom) boundary conditions on the rungs. }
\label{fig:gap_4xL}
\end{figure}
Finally, we consider the entanglement gap of 4-leg ladders.  We first consider the case where 
we split the system into two $2\times L$ ladders.  Here both subsystems are gapped and we might expect that the entanglement spectrum is also gapped regardless of the boundary conditions on the rungs. The numerical results shown in Fig.~\ref{fig:gap_4xL} confirm our expectations and we observe a tendency towards a finite gap for $L\to\infty$, both for open (top) and periodic (bottom) boundary conditions on the rungs. 
While the system sizes considered are insufficient to draw definitive conclusions, no qualitative difference between open and periodic boundary conditions can be observed from the plots.

For the $4\times L$ ladders we can obtain further insight by analyzing the entanglement Hamiltonian in the weak leg coupling limit.
We first consider the case of open boundary conditions on the rungs at $J_{\rm leg} =0$.
The ground state of the ladder in this case is 
\begin{equation}
\ket{\textrm{GS}} = \otimes_i \ket{s}_i
\end{equation}
where $\ket{s}_i$ is the lowest lying singlet state on a rung:
\begin{eqnarray}
\ket{s}_i &=& \alpha(\ket{\ua\ua\da\da}+\ket{\da\da\ua\ua})+\beta (\ket{\ua\da\da\ua} + \ket{\da\ua\ua\da})\cr\cr
&& +\gamma(\ket{\ua\da\ua\da}+\ket{\da\ua\da\ua} ),
\end{eqnarray}
and the parameters $\alpha, \beta$, and $\gamma$ are defined as
\begin{eqnarray}
\alpha &=& \frac{1}{\sqrt{12(2+\sqrt{3})}};\cr\cr 
\beta &=& \frac{2+2\sqrt{3}}{2 \sqrt{12(2+\sqrt{3})}};\cr\cr
\gamma &=& -\frac{4+2\sqrt{3}}{2 \sqrt{12(2+\sqrt{3})}}.
\end{eqnarray}
If we now perform a partial trace of sites $1$ and $2$ on each rung we obtain a reduced density matrix of the form
\begin{equation}
\rho^0_{\rm red} = \prod_i((4(\alpha^2-\frac{1}{4}){\bf S}_{3i}\cdot{\bf S}_{4i}+\frac{{\bf I}}{4})),
\end{equation}
which in turn implies an entanglement Hamiltonian given by
\begin{equation}
H^0_{\rm entanglement} = -\sum_i \log\big( 4(\alpha^2-\frac{1}{4}){\bf S}_{3i}\cdot{\bf S}_{4i}+\frac{{\bf I}}{4}\big).
\end{equation}
This implies the ground state of $H^0_{\rm entanglement}$ is a product of rung singlets with an excitation gap to a rung triplet
of $E_{\rm ent. gap} = -\log (\alpha^2/(1-3\alpha^2))$.

\begin{figure}
\centering
\includegraphics[width=0.9\columnwidth]{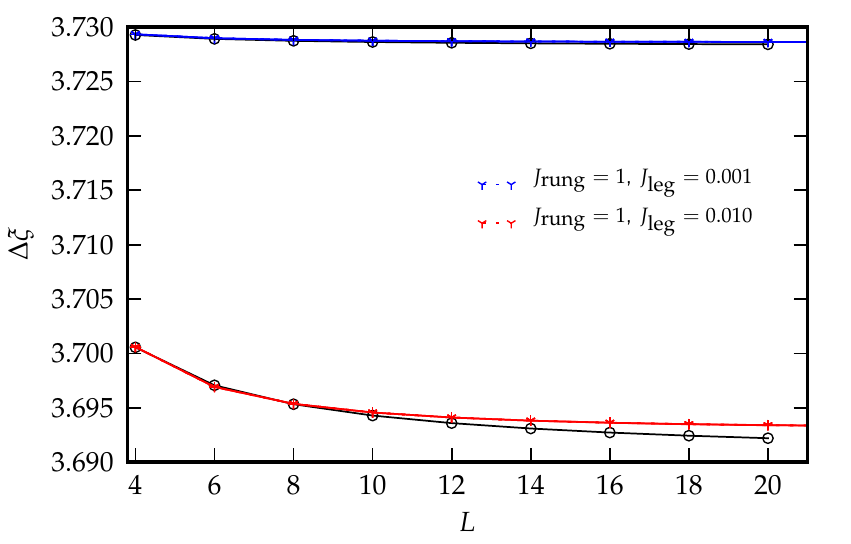}
\includegraphics[width=0.9\columnwidth]{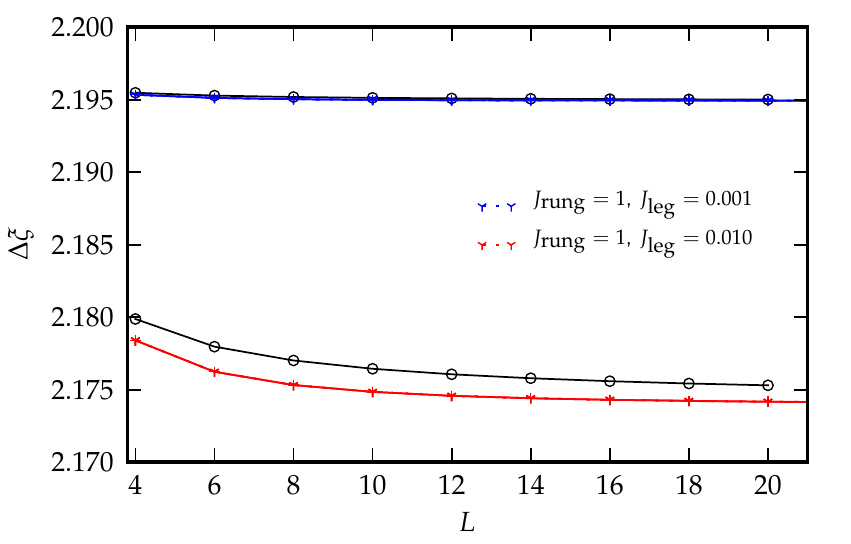}
\caption{Entanglement spectral gap for a bipartite splitting of a $4\times L$ ladder into two $2\times L$ ladders for the ground state of the ladder Heisenberg Hamiltonian with open (top) and periodic (bottom) boundary conditions on the rungs. The analytic computations are plotted with open (black) circles. }
\label{fig:gap_4xLX}
\end{figure}

We now consider the effects of the presence of a weak $J_{\rm leg}$.   In first order perturbation theory, the ground state
product of singlets is mixed in with various excited rung triplets (three in total).  The correction to the ground state
energy takes the form
\begin{eqnarray}\label{deltaGS}
\delta \ket{\rm GS} &=& \frac{J_{\rm leg}}{J_{\rm rung}} \sum_{\mu=1,2,3}c_\mu \sum_i \ket{s}_1\otimes \cdots \otimes \ket{s}_{i-1}\cr\cr &&\hskip -.25in 
\otimes (|t_\mu^+\rangle_i |t_\mu^-\rangle_{i+1}+|t_\mu^-\rangle_i |t_\mu^+\rangle_{i+1} + |t_\mu^0\rangle_i |t_\mu^0\rangle_{i+1})\cr\cr
&&\hskip -.25in \otimes \ket{s}_{i+2}\otimes \cdots \otimes \ket{s}_L,
\end{eqnarray}
where the coefficients $c_\mu$ and the states $|t^{+,-,0}_{1,2,3}\rangle$ are defined in the Appendix.
This correction to the ground state energy then leads to a correction to the reduced density matrix of
the form:
\begin{eqnarray}\label{deltarho}
\delta \rho  &=& \frac{J_{\rm leg}}{J_{\rm rung}}\sum_{i=1}^{N-1}\bigg[\prod^{i-1}_{j=1}(4(\alpha^2-\frac{1}{4}){\bf S}_{3j}\cdot {\bf S}_{4j}+\frac{I_j}{4})
\delta \rho_{i,i+1}\cr\cr
&&\hskip .5in \times\prod^{N}_{j=i+2}(4(\alpha^2-\frac{1}{4}){\bf S}_{3j}\cdot {\bf S}_{4j}+\frac{I_j}{4})\bigg];\cr\cr
\delta\rho_{i,i+1} &=& \bigg(J_{33}{\bf S}_{3i}\cdot {\bf S}_{3i+1} \cr\cr
&& \hskip -.25in + J_{34}{\bf S}_{3i}\cdot {\bf S}_{4i+1}
+  J_{43}{\bf S}_{4i}\cdot {\bf S}_{3i+1} + J_{44}{\bf S}_{4i}\cdot {\bf S}_{4i+1} \cr\cr
&&\hskip -.25in + J_{3344}{\bf S}_{3i}\cdot{\bf S}_{3i+1}{\bf S}_{4i}\cdot{\bf S}_{4i+1}\cr\cr
&&\hskip -.25in + J_{3443}{\bf S}_{3i}\cdot{\bf S}_{4i+1}{\bf S}_{4i}\cdot{\bf S}_{3i+1}\bigg),
\end{eqnarray}
where the $J$'s are given in the Appendix.
We see that $\delta\rho$ contains all possible couplings consistent with SU(2)
invariance between nearest neighbor rungs including a number of four spin terms.  Unlike the
two-leg ladder \cite{poilblanc,lauchli}, we will thus not obtain a particularly simple form for the entanglement
Hamiltonian.  The lowest entanglement excitation is a $k=\pi$ triplet, 
$$
\ket{\psi_{t^{+,-,0}}}(k=\pi) = \sum_i (-1)^i \ket{\psi_{t^{+,-,0}}}_i
$$
where
$$
\ket{\psi_{t^{+,-,0}}}_i \equiv \prod_{j=1}^{i-1}\ket{s}_j\otimes |t^{+,-,0}\rangle_i \otimes \prod_{j=i+1}^N \ket{s}_j .
$$
Here $|s\rangle$ and $|t^{+,-,0}\rangle$ are states on a two-site rung.
The entanglement gap is then
\begin{eqnarray}
E_{\rm ent. gap} &=& -\log \bigg(\frac{\alpha^2}{1-3\alpha^2} - \cr\cr
&& \hskip -.55in \frac{J_{\rm leg}}{J_{\rm rung}}\frac{L-1}{2L} \frac{J_{33} + J_{44}-J_{34}-J_{43}+J_{3344}}{(1-3\alpha^2)^2}\bigg).
\end{eqnarray} 
We note that this expression is only valid at relatively small $J_{\rm leg}$, see Fig.~\ref{fig:gap_4xLX} (top) for $J_{\rm leg} = 0.001, 0.01$.

We also consider the entanglement Hamiltonian for the four leg ladder with periodic boundary conditions.
At $J_{\rm leg} = 0$ we find
\begin{equation}
H^{0,\textrm{PBC}}_{\textrm{entanglement}} = -\sum_i \log\bigg(-\frac{2}{3}{\bf S}_{3i}\cdot{\bf S}_{4i}+\frac{{\bf I}}{4}\bigg).
\end{equation}
This leads to an entanglement gap (to a triplet) of $E_{\rm ent. gap} = \log 9$ for $J_{\rm leg} = 0$.

And again we will compute the correction at first order in $J_{\rm leg}$ to the entanglement gap.  For periodic boundary
conditions the correction to the ground state energy again involves mixing with the three possible rung triplet excitations
(as in Eqn. (\ref{deltaGS}).  Correspondingly the correction to the reduced density matrix has the same form as in Eqn. (\ref{deltarho}).  We give some of the details of this computation in the Appendix.  The entanglement
excitation with minimal gap for this case is also a $k=\pi$ triplet.  Its gap is equal to
\begin{eqnarray}
E^{\rm PBC}_{\rm ent. gap} &=& -\log \bigg(\frac{1}{9} -\frac{J_{\rm leg}}{J_{\rm rung}}\frac{L-1}{L}\frac{1}{2}\big(\frac{4}{3}\big)^2\cr\cr
&& \hskip -.7in \times(J^{\rm PBC}_{33} + J^{\rm PBC}_{44}-J^{\rm PBC}_{34}-J^{\rm PBC}_{43}+J^{\rm PBC}_{3344})\bigg).
\end{eqnarray} 
Again this is only valid for small $J_{\rm leg}$, see Fig.~\ref{fig:gap_4xLX} (bottom).

\begin{figure}
\centering
\includegraphics[width=0.9\columnwidth]{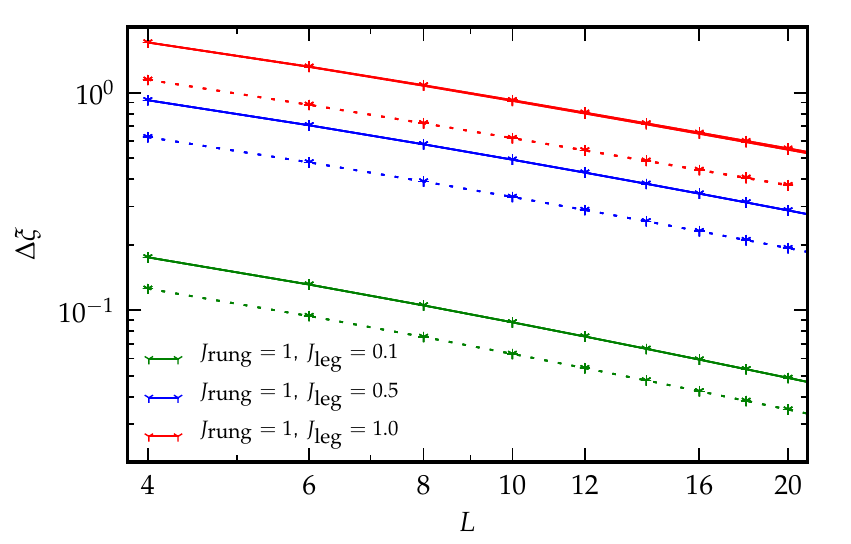}
\caption{Entanglement spectral gap for a bipartite splitting of a $4\times L$ ladder into a chain of length $L$ and $3 \times L$ ladders for the ground state of the ladder Heisenberg Hamiltonian with open (solid lines) and periodic (dashed lines) boundary conditions on the rungs. }
\label{fig:gap_4xL13}
\end{figure}
{
As our final result, we consider splitting the $4\times L$ ladder into a chain and a $3\times L$ ladder. In this case, both of the subsystems are gapless and we expect a gapless entanglement spectrum. The numerical results shown in Fig.~\ref{fig:gap_4xL13} confirm this hypothesis.  Unlike
the case of three-leg ladders, the boundary conditions on
the four-leg ladder do not play a significant role here.

From the computational point of view, the simulation of the excited states for the entanglement Hamiltonian
of the four leg ladders requires at most an hour. However, obtaining the reduced density matrix as a manageable MPO requires substantial memory resources (several gigabytes) as the matrix product operator obtained by simply contracting the tensor network for the ground state (Fig.~\ref{fig:rhompo}) is very large (in a redundant way) and must be aggressively truncated, requiring up to an hour of processor time for each case. 

The spectral gap (but not other levels in the entanglement spectrum) could alternatively have been 
simulated by the simulating separately the ``ground state'' of the matrix product operator $\rho_S$ 
in the $S^z=0$ and in the $S^z=1$ sector.

\section{Conclusions}
We have described a method to simulate quantum many-body systems using tree tensor networks where the global symmetries are employed to reduce the minimization costs of the tensors to a minimum by optimizing each charge configuration in the tensor individually. 
Furthermore, we have presented a method to calculate the entanglement spectrum for
large many-body systems which can be described in terms of tensor networks by 
by constructing the reduced density operator as a matrix product operator and calculating eigenvectors in various symmetry sectors. The method can be used on top of our tree tensor network method to simulate the ground states but also on top of the standard DMRG algorithm. 

We have used the methods described in this manuscript to simulate the entanglement spectra of $2\times L$, $3\times L$ and $4\times L$ ladders with either open or periodic boundary conditions on the rung. From numerical results we have found that the nature of the entanglement spectrum depends not only on the subsystem in the bipartite splitting but also on the number of boundaries connecting the systems, as a result of the boundary conditions in the Hamiltonian operator.  Unlike the case of two-leg ladders, in the limit of weak coupling along
the legs of the ladder, we in general did not find a simple relationship between the entanglement Hamiltonian and the minimal Heisenberg Hamiltonian of the untraced subsystem.

\begin{acknowledgments}
IP  thanks RMK and Brookhaven National Laboratory for hospitality during his stay there where this work was initiated. This work was supported through the National Competence Center in Research (NCCR) QSIT,
the EU project QUEVADIS, the FWF SFB project ViCoM, and by the US DOE under contract number
DE-AC02-98 CH 10886.
The simulations were run on the Brutus cluster at ETH Zurich and on Vienna Scientific Cluster.
\end{acknowledgments}

\hskip .5in
\appendix

\section{Computational Details for the ${\cal O}(J_{\rm leg})$ Correction to the Bipartite Reduced Density Matrix, $\rho$, for the Four Leg Ladder}

\subsection{Open Boundary Conditions}
The excited rung triplets of the four site rung have the form:
\begin{widetext}
\begin{eqnarray}
|t^-_1\rangle &=&  -|\da\da\da\ua\rangle + (1+\sqrt{2})|\da\da\ua\da\rangle  - (1+\sqrt{2})|\da\ua\da\da\rangle + |\ua\da\da\da\rangle; \cr\cr
|t^+_1\rangle &=&  -|\ua\ua\ua\da\rangle + (1+\sqrt{2})|\ua\ua\da\ua\rangle  - (1+\sqrt{2})|\ua\da\ua\ua\rangle +
|\da\ua\ua\ua\rangle; \cr\cr
|t^0_1\rangle &=&  -|\ua\ua\da\da\rangle + (1+\sqrt{2})|\ua\da\ua\da\rangle  - (1+\sqrt{2})|\da\ua\da\ua\rangle +
|\da\da\ua\ua\rangle; \cr\cr
|t^-_2\rangle &=&  |\da\da\da\ua\rangle - |\da\da\ua\da\rangle  -|\da\ua\da\da\rangle + |\ua\da\da\da\rangle ;\cr\cr
|t^+_2\rangle &=&  |\ua\ua\ua\da\rangle - |\ua\ua\da\ua\rangle  - |\ua\da\ua\ua\rangle +
|\da\ua\ua\ua\rangle; \cr\cr
|t^0_2\rangle &=&  -|\da\ua\ua\da\rangle + |\ua\da\da\ua\rangle; \cr\cr
|t^-_3\rangle &=&  -|\da\ua\ua\da\rangle + |\ua\da\da\ua\rangle  + (-1+\sqrt{2})|\da\ua\da\da\rangle + |\ua\da\da\da\rangle; \cr\cr
|t^+_3\rangle &=&  -|\ua\ua\ua\da\rangle + (1-\sqrt{2})|\ua\ua\da\ua\rangle  + (-1+\sqrt{2})|\ua\da\ua\ua\rangle +
|\da\ua\ua\ua\rangle; \cr\cr
|t^0_3\rangle &=&  -|\ua\ua\da\da\rangle + (1-\sqrt{2})|\ua\da\ua\da\rangle  + (-1+\sqrt{2})|\da\ua\da\ua\rangle +
|\da\da\ua\ua\rangle .
\end{eqnarray}
Their energies are respectively $E_{t1}=\frac{1}{4}(-1-2\sqrt{2}), E_{t2} = -1/4$, and $E_{t3}= \frac{1}{4}(-1+2\sqrt{2})$.
The coefficients, $c_{\mu}$, that determine how these states contribute to $\delta \ket{\rm GS}$ (see Eqn. 14) are as follows:
\begin{eqnarray}
c_1 &=& \frac{11+7\sqrt{2}+6\sqrt{3}+4\sqrt{6}}{6(2+\sqrt{2})(2+\sqrt{3})}\frac{1}{\frac{1}{2}(-2-2\sqrt{3}+2\sqrt{2})};\cr\cr
c_2 &=& \frac{-11+7\sqrt{2}-6\sqrt{3}+4\sqrt{6}}{6(-2+\sqrt{2})(2+\sqrt{3})}\frac{1}{\frac{1}{2}(-2-2\sqrt{3})};\cr\cr
c_3 &=& \frac{1}{3}\frac{1}{\frac{1}{2}(-2-2\sqrt{3}-2\sqrt{2})}.
\end{eqnarray}
Finally the couplings defining $\delta\rho$ in Eqn. (\ref{deltarho}) at ${\cal O}(J_{\rm leg})$ are given by:
\begin{eqnarray}
J_{33} &=& c_1\frac{9+5\sqrt{2}+5\sqrt{3}+3\sqrt{6}}{12(2+\sqrt{2})(2+\sqrt{3})}+ \frac{c_2}{6} + c_3\frac{-9+5\sqrt{2}-5\sqrt{3}+3\sqrt{6}}{12(-2+\sqrt{2})(2+\sqrt{3})};\cr\cr
J_{44} &=& c_1\frac{13+9\sqrt{2}+7\sqrt{3}+5\sqrt{6}}{12(2+\sqrt{2})(2+\sqrt{3})}+ \frac{c_2}{6} + c_3\frac{-13+9\sqrt{2}-7\sqrt{3}+5\sqrt{6}}{12(-2+\sqrt{2})(2+\sqrt{3})};\cr\cr
J_{34} &=& J_{43} = -c_1\frac{10+7\sqrt{2}+6\sqrt{3}+4\sqrt{6}}{12(2+\sqrt{2})(2+\sqrt{3})} - \frac{c_2}{6} + c_3\frac{10-7\sqrt{2}+6\sqrt{3}-4\sqrt{6}}{12(-2+\sqrt{2})(2+\sqrt{3})};\cr\cr
J_{3344} &=& -J_{3443} = -c_1\frac{1}{6}(2+\sqrt{2}) -c_2\frac{2+\sqrt{3}}{3}+c_3\frac{3-2\sqrt{2}}{3(-2+\sqrt{2})}.
\end{eqnarray}
\end{widetext}

\subsection{Periodic Boundary Conditions}
We now present a similar set of data for the case of periodic boundary conditions.  The excited triplets on the 
four site rung are as follows:
\begin{widetext}
\begin{eqnarray}
|t^-_1\rangle &=&  -|\da\da\da\ua\rangle + |\da\da\ua\da\rangle  - |\da\ua\da\da\rangle + |\ua\da\da\da\rangle; \cr\cr
|t^+_1\rangle &=&  -|\ua\ua\ua\da\rangle + |\ua\ua\da\ua\rangle  - |\ua\da\ua\ua\rangle +
|\da\ua\ua\ua\rangle; \cr\cr
|t^0_1\rangle &=&  -|\ua\da\ua\da\rangle + |\da\ua\da\ua\rangle; \cr\cr
|t^-_2\rangle &=&  -|\da\da\da\ua\rangle - |\da\da\ua\da\rangle  + |\da\ua\da\da\rangle + |\ua\da\da\da\rangle; \cr\cr
|t^+_2\rangle &=& -|\ua\ua\ua\da\rangle - |\ua\ua\da\ua\rangle  + |\ua\da\ua\ua\rangle +
|\da\ua\ua\ua\rangle; \cr\cr
|t^0_2\rangle &=&  -|\ua\ua\da\da\rangle + |\da\da\ua\ua\rangle; \cr\cr
|t^-_3\rangle &=& |\da\da\da\ua\rangle - |\da\da\ua\da\rangle  - |\da\ua\da\da\rangle + |\ua\da\da\da\rangle;
\cr\cr
|t^+_3\rangle &=&  |\ua\ua\ua\da\rangle - |\ua\ua\da\ua\rangle - |\ua\da\ua\ua\rangle +|\da\ua\ua\ua\rangle; \cr\cr
|t^0_3\rangle &=&  -|\da\ua\ua\da\rangle + |\ua\da\da\ua\rangle;
\end{eqnarray}
Their energies are respectively $E_{t1}=-1, E_{t2} = 0$, and $E_{t3} = 0$.
The coefficients, $c_{\mu}$, that determine how these states contribute to $\delta|GS\rangle$ are equal to:
\begin{eqnarray}
c^{\rm PBC}_1 &=& -\frac{1}{3};\cr\cr
c^{\rm PBC}_2 &=& -\frac{1}{24};\cr\cr
c^{\rm PBC}_3 &=& -\frac{1}{24}.
\end{eqnarray}
Finally the couplings defining $\delta\rho$ in Eqn. (\ref{deltarho}) at ${\cal O}(J_{\rm leg})$ are given by
\begin{eqnarray}
J^{\rm PBC}_{33} &=& J^{\rm PBC}_{44} = \frac{c_1}{3}+ \frac{c_2}{12} + \frac{c_3}{12};\cr\cr
J^{\rm PBC}_{34} &=& J^{\rm PBC}_{43} = -\frac{c_1}{3}+ \frac{c_2}{12} - \frac{c_3}{12};\cr\cr
J^{\rm PBC}_{3344} &=& -J^{\rm PBC}_{3443} = -\frac{c_1}{3} - \frac{4c_3}{3}.
\end{eqnarray}
\end{widetext}

\end{document}